\documentclass[aps,pra,showpacs,twoside,twocolumn,10pt]{revtex4-2}
\usepackage[colorlinks=true, citecolor=blue, urlcolor=blue, linkcolor = blue ]{hyperref}
\usepackage{epsfig,newlfont,amssymb,amsfonts,amsmath,bm,subfigure,palatino,mathtools,amsthm,braket,times,soul,enumitem,color}
\usepackage[normalem]{ulem}
\newcommand{\stkout}[1]{\ifmmode\text{\sout{\ensuremath{#1}}}\else\sout{#1}\fi}
\usepackage[T1]{fontenc}
\usepackage[normalem]{ulem}
\usepackage{graphicx}
\usepackage{placeins}
\usepackage{hyperref}
\usepackage{multirow}
\usepackage{hhline,booktabs}

\begin{document}
\title{{Estimating phase transition of perturbed $J_1-J_2$ Heisenberg quantum chain \\ in mixtures of ground and first excited states}}
\author{Sayan Mondal$^1$, George Biswas$^2$, Ahana Ghoshal$^1$, Anindya Biswas$^{2*}$, Ujjwal Sen$^{1*}$}
\affiliation{$^1$Harish-Chandra Research Institute, A CI of Homi Bhabha National Institute, Chhatnag Road, Jhunsi, Prayagraj 211 019, India\\
$^2$National Institute of Technology Sikkim, Ravangla, South Sikkim 737 139, India\\
\(^*\)Correspondence can be sent to anindya@nitsikkim.ac.in and ujjwal@hri.res.in.}

\begin{abstract}
We show that the nearest neighbour entanglement in a mixture of ground and first excited states - a subjacent state - of the $J_1-J_2$ Heisenberg quantum spin chain can be used as an order parameter to detect the phase transition of the chain from a gapless spin fluid to a gapped dimer phase. We study the effectiveness of the order parameter for varying relative mixing probabilities between the ground and first excited states in the subjacent state for different system sizes, and extrapolate the results to the thermodynamic limit. We observe that the nearest neighbour concurrence can play a role of a good order parameter even if the system is in the ground state, but with a small finite probability of leaking into the first excited state. Moreover, we apply the order parameter of the subjacent state to investigate the response to separate introductions of anisotropy and of glassy disorder on the phase diagram of the model, and analyse the corresponding finite-size scale exponents and the emergent tricritical point in the former case. The anisotropic $J_1-J_2$ chain has a richer phase diagram which is also clearly visible by using the same order parameter.   
\end{abstract} 

\maketitle
  
\section{Introduction}
The study of quantum many-body systems utilizing the concepts of quantum information theory is a fruitful avenue for investigating various many-body quantum phenomena like quantum phase transitions~\cite{sachdev_2011,doi:10.1080/00018730701223200,RevModPhys.80.517,Carr2011,Lewenstein2012}. Investigations in this direction include Refs.  \cite{PhysRevA.66.032110,Osterloh2002}, wherein bipartite quantum entanglement~\cite{Horodecki2009,GUHNE2009,Das2019} was used as an order parameter for detection of the quantum phase transition in the one-dimensional transverse Ising and other XY quantum spin models. Later on, multipartite quantum entanglement measures like geometric~\cite{shimony1995degree,HBarnum_2001,Blasone2008,Das2019} and generalized geometric measure ~\cite{Das2016,SenDe2010}, and other quantum correlation measures like quantum discord~\cite{Henderson_2001,Ollivier2001}, have been shown to be effective detectors of quantum phase transitions in many-body quantum systems.
See e.g. Refs.~\cite{quant-ph/0312154,PhysRevA.71.053804,PhysRevLett.94.147208,PhysRevA.72.022314,PhysRevLett.94.163601,PhysRevLett.95.056402,PhysRevA.73.010305,Cavalcanti_2006,PhysRevA.74.022314,Oliveira2006,PhysRevB.75.165106,Costantini_2007,Dahlsten_2007,Sen2010,Bera2012,Biswas2014} and  Refs.~\cite{doi:10.1142/S123016121440006X,Bera_2018} in this regard. A strong connection between entanglement and many body physics has been established by the numerical simulation of quantum many-body systems using matrix product states~\cite{Klumper_1991,Klumper_1992,Garcia2007}, projected entangled pair states~\cite{Verstraete2004},  density matrix renormalization group~\cite{White1992,White1993,Vidal2003,Verstraete2004,Karen2006}, tensor network states~\cite{Niggemann1997,Verstraete20041,Levin2007,Jiang2008,Jordan2008,Gu2008}, etc.  
Experimental evidence of  these and related concepts in quantum many-body systems are found in atoms in an optical lattice~\cite{Bloch2008,Nayak2008}, trapped ions~\cite{Leibfried2003,HAFFNER2008155}, photons
~\cite{Aspuru-Guzik2012}, and nuclear magnetic resonance systems
~\cite{RevModPhys.76.1037} of particles. Furthermore, the connection between quantum information theory and many-body quantum systems helps in quantum state engineering 
which facilitates the development of realistic quantum computation.

The phenomenon of quantum phase transition in the $J_1-J_2$ Heisenberg spin model, where $J_1$ and $J_2$ stand for the nearest and next to nearest neighbour coupling constants respectively,  is a well known one. See e.g. ~\cite{Majumdar1969,White1996,Gu2004,Mikeska2004,Chhajlany2007}. 
Classically, phase transitions occur when the system approaches a certain critical temperature, beyond which the macroscopic properties of the system change~\cite{Goldenfeld1992,Yeomans1992,Onuki2002,Ma2019}. Quantum phase transitions take place at absolute zero temperature, when some external parameter or coupling strength is varied~\cite{sachdev_2011,Carr2011,Dutta2010}. The quantum phase transition of the one-dimensional antiferromagnetic $J_1-J_2$ model
from a gapless fluid to gapped dimer phase 
was investigated in~\cite{Haldane1982,Tonegawa1987,OKAMOTO1992433,Xu2021,PhysRevB.54.R9612}
using exact digonalization and field theory formalism. Subsequently, measures of bipartite entanglement, like concurrence~\cite{PhysRevLett.78.5022,PhysRevLett.80.2245}, quantum
fidelity~\cite{Liang_2019}, and valence bond entanglement~\cite{Alet2010} have been shown to be good detectors of the quantum phase transition, considering the respective quantities in
the first excited state, but not in the ground state~\cite{Biswas_2020,Chen2007}. The
measure of multipartite entanglement, generalized geometric
measure, also could not indicate the quantum phase transition
while considering the ground state~\cite{Biswas2014}.

Quantum systems are usually prone to noise from the environment. One of the most common forms of noise is thermal noise, and a many-body system prepared at very low temperatures, is practically a mixture of its low-lying energy states. The study of critical phenomena and quantum information measures of such systems are extensively done for the ground state. It is however interesting to  study other low-lying states as well, as the system can only be in the ground state at the absolute zero temperature, which is not practical. 

In this paper, we show that the nearest neighbour concurrence of a mixture of ground and first excited states of the one-dimensional quantum Heisenberg $J_1-J_2$ model can behave as a good indicator of the quantum phase transition in the model. We consider a mixture of the ground and first excited states to mimic the state of the system at some finite temperature - the higher excited states, being present with lower probabilities, are ignored. We find that the two-party entanglement of such a state indicates the presence of the critical point in the ground state of the model. We are therefore trying to find ``shadows'' of the ground state quantum phase transitions in a given physical system, in states of the same system which have a finite but low temperature. This is physically meaningful, as a real state of a system will always be of the latter type. Although the concept of using a mixture of ground and first excited states and identifying entanglement as a good order parameter of phase transition in this model are already studied in the literature, in this paper we try to investigate this matter more deeply.
Recently, concurrence~\cite{PhysRevLett.78.5022,PhysRevLett.80.2245} and shared purity~\cite{Biswas12014}, have been shown to be good quantum phase transition indicators in a mixture of ground and low-lying energy states for the $J_1-J_2$ model on a spin chain~\cite{Biswas2022}. In that work, they have considered the mixture of ground and low-lying energy states with certain mixing probabilities. On the contrary, in this paper we find that for arbitrarily small mixing probabilities also, the critical point can be observed. 
Furthermore, we investigate the effect of mixing probabilities on the phase transition points for different system-sizes and a finite-size scaling analysis is done to find the critical point in the thermodynamic limit. Moreover, we analyse the effect of introduction of an anisotropy parameter in the nearest neighbour and next-nearest neighbour coupling constants of the system and in particular investigate the change in the phase transition point on varying the anisotropy parameter. A scaling analysis is done here as well. A phase diagram of the system is clearly visible for this case. From the depictions given in this paper, we can see the phase boundaries among the various phases of the system like spin fluid state, dimer phase and N\'{e}el phase. In previous works, the phase boundaries were obtained using various methods like level-spectroscopy \cite{10.1143/PTPS.145.113} and tensor network methods \cite{Sato2011COMPETINGPI}. In this work, we obtain a similar phase boundary, using nearest-neighbour concurrence. 
We reiterate here that quantum phase transitions happen only at zero temperature, and so the studies of finite low-temperature states or low-lying eigenstates detect the traces of zero-temperature transitions that are still visible in a state that is not a zero-temperature state. 

In practical scenarios, it is neither possible to reach absolute zero-temperature nor is it possible to completely isolate our system from the environment. This makes it quite difficult to fabricate a lattice chain with $N$ quantum spin$-\frac{1}{2}$ particles. There may arise fluctuations in the positions of the spins in the lattice or in the tuning parameters of the set-up, or the magnetic field may not be homogeneous. These imperfections are very difficult to keep track of and are usually modelled as disorder.  It is therefore both natural and potentially useful to study the effect of disorder in our system. Sometimes, the presence of disorder may provide some advantages over the ideal situation ~\cite{Anderson1958, Brout1959, Aharony1978, Villain1980, KURMANN1982, Anderson1984, Lee1985, Binder1986, Chowdhury1986, Mezard1988, Henley1989, Barabasi1999, Goltsev2003, Ahufinger2005, Volovik2006, de2006random, Adamska2007, Abanin2007, Das2008, FALLANI2008, Niederberger2008, aspect2009anderson, Alloul2009,Hide2009,Niederberger_2009,Sanchez-Palencia2010, Modugno_2010, Niederberger2010,Tsomokos2011,auerbach2012interacting,Shapiro_2012,Hide_2012,Zuniga2013,Foster2014,Martin2014,Majdandzic2014,Kjall2014,Yao2014,Eisert2015,Cakmak2015,Sadhukhan12016,Sadhukhan2016,Bera2016,Rakshit2017,stauffer2018,Bera2019,Ghosh2020,Ghoshal2020,Sarkar2022}. This type of disordered system is more realistic and we show that entanglement still behaves as an identifier of phase transition even if the system is some distance away from its ideal disorder-free nature. In our work, we introduce a disorder parameter in the system and obtain the shifts in the critical point of the system in the presence of glassy Gaussian disorder.  

The remainder of the paper is arranged as follows. The one-dimensional $J_1-J_2$ Heisenberg spin model and the definition of the quantum information order parameter, concurrence, is discussed in Sec.~\ref{Sec:2}. The behaviour of nearest neighbour concurrence, especially its dependence on the mixing probability of ground and first excited states is presented in Sec.~\ref{Sec:3}. A finite-size scaling for the extrapolation of the results in the thermodynamic limit is also presented in this section. In Sec.~\ref{Sec:4}, we introduce an anisotropy parameter in the system and note the shift of the phase transition point depending on the parameter and extrapolate the result to the thermodynamic limit. Section~\ref{Sec:5} contains the behaviour of nearest neighbour concurrence in the presence of glassy Gaussian disorder. The concluding remarks are presented in Sec.~\ref{Sec:6}. 

\begin{figure*}
\centering
\includegraphics[width=8.7cm]{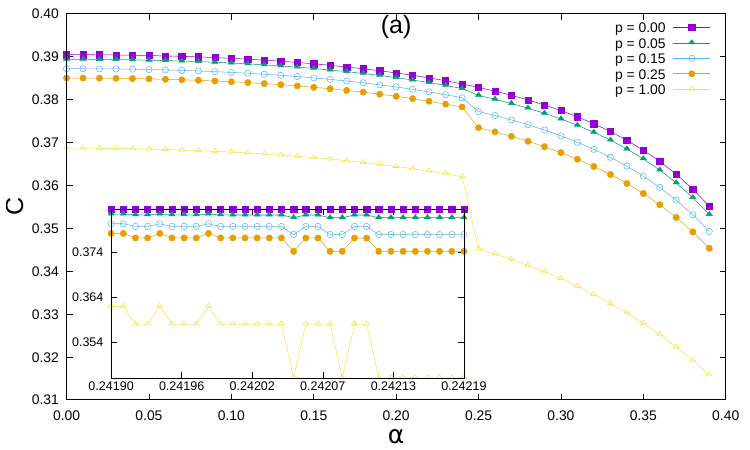}%
\hspace{.25cm}%
\includegraphics[width=8.7cm]{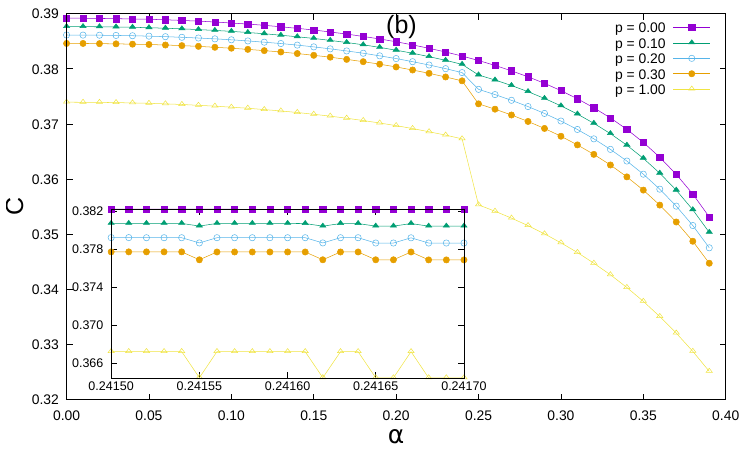}
\caption{Nearest neighbour concurrence of the $J_1-J_2$ model. The concurrence between two consecutive spin particles is plotted with respect to the relative coupling strength $\alpha$ for different values of $p$ of the subjacent state for (a) $N = 20$ and (b) $N=24$. The regions near the transition points are magnified and shown in the respective insets. The different colors of the lines with the different symbols represent the different values of $p$, which are mentioned in the legends. The quantity plotted along the horizontal axes is dimensionless, whereas the same along the vertical axes are in ebits.}
\label{fig:N20_1}
\end{figure*}

\section{Motivating the choice of system state}
\label{Sec:2}
The one-dimensional $J_1-J_2$ Heisenberg spin model, having the nearest and next-nearest neighbour coupling constants as $J_1$ and $J_2$ respectively, 
can be described by the Hamiltonian,
\begin{equation}\label{eq:H}
H = J_1 \hbar \sum_{i = 1}^N  {\vec{\sigma}_i\cdot\vec{\sigma}_{i+1}} + J_2 \hbar \sum_{i = 1}^N  {\vec{\sigma}_i\cdot\vec{\sigma}_{i+2}},
\end{equation}
with $N$ being the number of spins of the system and $\vec{\sigma} = \sigma^x \hat{x} + \sigma^y \hat{y} + \sigma^z \hat{z}$, where $\sigma^x$, $\sigma^y$ and $\sigma^z$ represent the Pauli matrices. \textcolor{black}{Here we employ periodic boundary conditions, which means that  $\vec{\sigma}_{N+1} = \vec{\sigma}_{1}$.} Here $J_1\hbar$ and $J_2\hbar$, having  the unit of energy, stand for the anti-ferromagnetic coupling constants 
and are therefore zero or positive real numbers. This model can describe some solid state systems like SrCuO$_2$~\cite{MATSUDA19951671}. {This model is also known as the Majumdar–Ghosh model for $\alpha=\frac{J_2}{J_1}=\frac{1}{2}$~\cite{Majumdar1969}. At $\alpha=\frac{1}{2}$, the model is exactly solvable and the ground state is doubly degenerate.}
 {The quantum phase transition of the one-dimensional antiferromagnetic $J_1-J_2$ model from a gapless fluid to gapped dimer phase driven by the change of the system  parameter $\alpha$ was investigated in~\cite{Haldane1982,Tonegawa1987,OKAMOTO1992433,Xu2021} using exact digonalisation and field theory formalism. It was found that the phase transition occurs around a critical value of the parameter, $\alpha_c \approx 0.241$.}
 {The anisotropic $J_1-J_2$ Heisenberg quantum spin model with an anisotropic coupling constant $\delta$ along the $z$ direction was studied in~\cite{Cloizeaux1966,Affleck1988,Hirata2000,Somma2001}. When $\alpha \lesssim 0.24$, for $\delta <1$, the system is in the spin fluid phase and for $\delta>1$, it goes into the N\'{e}el phase. In the spin fluid phase, no ordered arrangement of spins is formed, and they remain fluctuating even at absolute zero temperature~\cite{doi:10.1126/science.abi8794, Valero2021}. N\'{e}el order phase is an antiferromagnetic phase below a sufficiently low temperature (known as N\'{e}el temperature)~\cite{PhysRevB.86.144411, refId0}. The phase diagram of this model in the range $0 \le \alpha \le \frac{1}{2}$ for $\delta \ge 0$ and $\delta\le 0$ is studied in~\cite{Nomura1994,Hirata2000}. The anisotropy constant $\delta$ plays an important role in the phase diagram of this model between $0 \le \alpha \le \frac{1}{2}$, and so it is important to study the effect of this anisotropy term on the "isotropic" phase transition from gapless fluid to gapped dimer phase around $\alpha_c \approx 0.241$. }

In the literature, the phase transition point was investigated
by determining the difference between the singlet-triplet and the singlet-singlet energy gaps for finite-size systems~\cite{Tonegawa1987,OKAMOTO1992433}. The singlet-triplet and singlet-singlet energy gaps are respectively given by
\begin{eqnarray*}
G_{st}(N,\alpha)&=& E_1^{(0)}(N,\alpha)-E_0^{(0)}(N,\alpha),\\
G_{ss}(N,\alpha)&=& E_0^{(1)}(N,\alpha)-E_0^{(0)}(N,\alpha).
\end{eqnarray*}
$E_m^{(0)}(N,\alpha)$ and $E_m^{(l)}(N,\alpha)$ are respectively the ground and the $l^{th}$ excited state energies in the total spin angular momentum, $S_{total}=m$ subspace. 
 {\textcolor{black}{The measures of bipartite entanglement, like concurrence~\cite{PhysRevLett.78.5022,PhysRevLett.80.2245}, and quantum fidelty~\cite{Liang_2019} have been shown to be good detectors of the quantum phase transition, considering the respective quantities in the first excited state, but not in the ground state~\cite{Biswas_2020,Chen2007}. So, while the phase transition at $\alpha \approx 0.2411$ is a zero-temperature phenomenon at the ground state, there could exist order parameters that find their shadows in the first excited state instead of the ground state. } The measure of multipartite entanglement, generalized geometric measure, also could not indicate the quantum phase transition while considering the ground state~\cite{Biswas2014}. } Thus, concurrence, a measure of two-qubit entanglement, can be considered as an order parameter for detecting the phase transition of the $J_1-J_2$ model, not in the ground state but in the first excited state~\cite{Biswas_2020} as well as in a mixture of ground and some low-lying energy states~\cite{Biswas2022}.

For a two-qubit density matrix, the concurrence $C$ 
is defined as
\begin{equation}
    C=\max(0, \sqrt{\lambda_1}- \sqrt{\lambda_2} - \sqrt{\lambda_3} - \sqrt{\lambda_4}),
\end{equation} 
where $\lambda_i$'s are the eigenvalues of $\rho\tilde{\rho}$ with $\lambda_1$ being the largest. Here, $\tilde{\rho}=(\sigma_y \otimes \sigma_y)\rho^*(\sigma_y \otimes \sigma_y)$ and $\rho^*$ is the complex conjugate of $\rho$ in the computational basis. 
In
this paper, we consider mixtures of the ground and first excited states and investigate whether its concurrence acts as a good order parameter. We consider a situation where there is a possibility of the spins to be in the first excited state of the Hamiltonian with probability $p$ and the ground state of the same with probability $(1-p)$. Here we take $p$ as the probability corresponding to the Maxwell-Boltzmann distribution, viz. $p=\frac{e^{-\frac{E^{(1)}(N,\alpha)-E^{(0)}(N,\alpha)}{k_BT}}}{1+e^{-\frac{E^{(1)}(N,\alpha)-E^{(0)}(N,\alpha)}{k_BT}}}$, where 
$k_B$ is the Boltzmann constant and $T$ is the absolute temperature of the system. {Here we have omitted the subscipts $m$ of $E$'s, as all energies are considered in the calculations.} 
A periodic boundary condition is imposed on the system, i.e., $\sigma_{N+1}=\sigma_1$. The quantum state 
being considered can therefore be represented as 
\begin{equation} \label{eq:state}
\rho = (1-p)|\Psi_0\rangle \langle\Psi_0| + \frac{p}{d} \sum_{i=1}^d |\Psi_{1}^{i}\rangle \langle\Psi_{1}^{i}|, 
\end{equation}
where $|\Psi_0\rangle$ is the ground state and $|\Psi_{1}^{i}\rangle$s are the first excited states with $d$-fold degeneracy. 
We will refer to this as the subjacent state. We have only considered systems with even number of spins, for this the ground state has no degeneracy. The motivation behind considering the subjacent state is that a physical system at non-zero temperature, will be in a thermal mixture of the eigenstates of the system and for very low temperatures, the ground state and first excited states will have considerably higher probability than the other excited states. In fact, for every small $p$, we can find a non zero finite temperature which corresponds to the probability distribution between the first excited state and ground state. In this paper we are restricting ourselves to small values of $p$, vis-\'a-vis low but non-zero temperatures, as for higher temperatures, higher energy states will also contribute which we are ignoring. For the numerical analysis, the ground and first excited states of the system are obtained by using the Lanczos method~\cite{Lanczos1950} of exact diagonalisation. The concurrence we obtain in the paper is the concurrence of two consecutive spins. Due to the symmetry of the Hamiltonian, and as we are imposing periodic boundary condition, the entanglement between any two nearest neighbour spins will be the same. In the succeeding sections, "concurrence" refers to the nearest neighbour concurrence. In the succeeding section, we investigate how the behaviour of concurrence as an order parameter depends on the values of $p$, for the $J_1-J_2$ model.


\begin{figure*} 
\includegraphics[width=8.7cm]{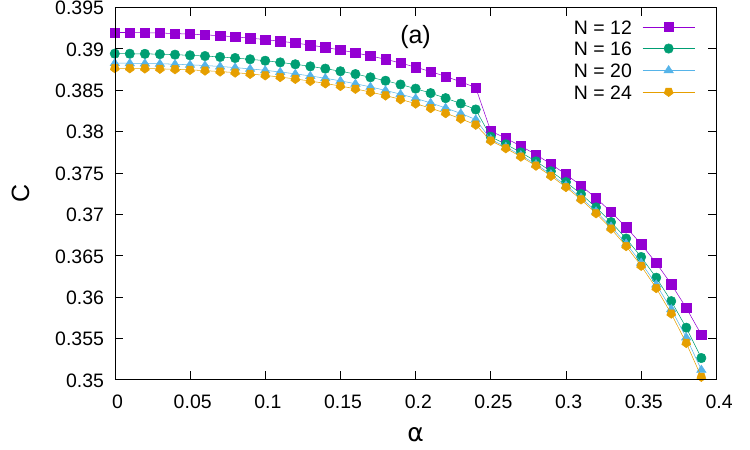}
\includegraphics[width=8.7cm]{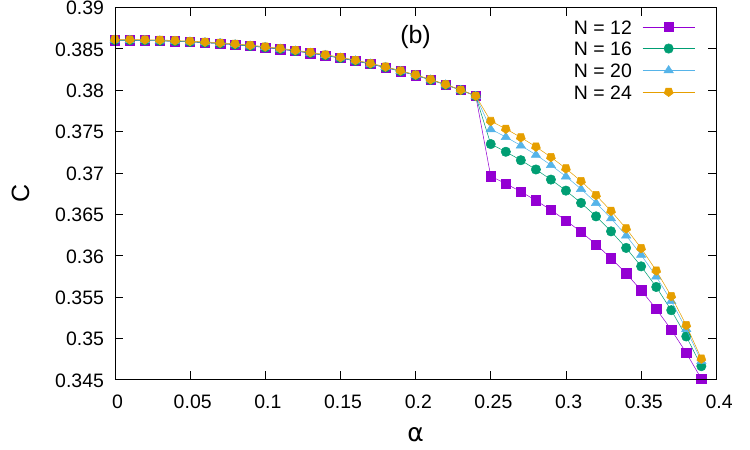}
\includegraphics[width=8.7cm]{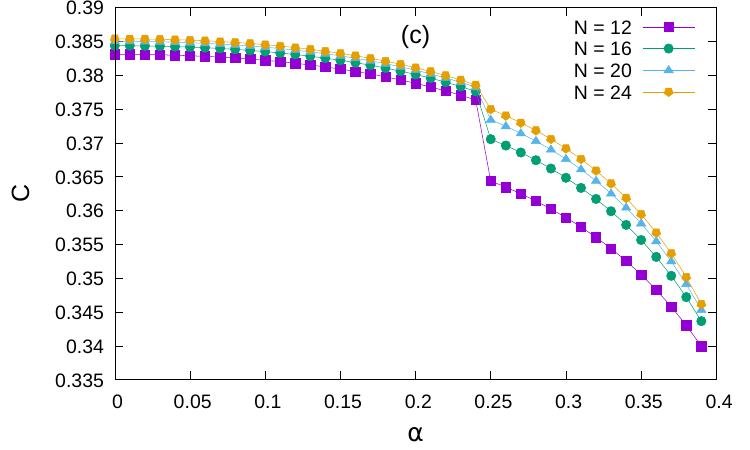}
\includegraphics[width=8.7cm]{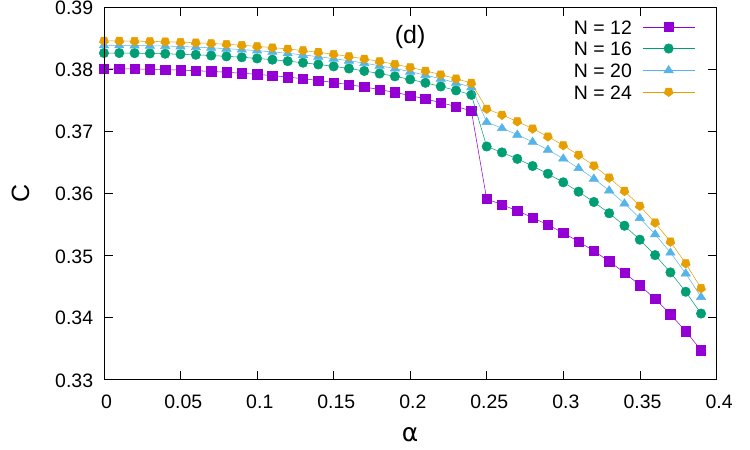}
\caption{Nearest neighbour concurrence of the $J_1-J_2$ model for different system sizes. Here we have depicted the nearest neighbour concurrence with relative coupling $\alpha$ for different values of $N$, for (a) $p = 0.10$, (b) $p = 0.20$, (c) $p = 0.25$ and (d) $p = 0.30$ of the subjacent state. The different colors of the lines with different symbols represent different values of $N$, which are mentioned in the legends. The quantity plotted along the horizontal axes is dimensionless and the same along the vertical axes is in ebits.}
\label{fig:1p20}
\end{figure*}

\section{Concurrence of subjacent state as order parameter in $J_1-J_2$ model}
\label{Sec:3}
The behaviour of concurrence as an order parameter 
strongly depends on the mixing probability $p$ of the ground and the first excited states.
In Fig. \ref{fig:N20_1}, we depict the nature of concurrence with the relative coupling strength $\alpha$ for different values of mixing probability $p$ for $N=20$ and $N=24$, in the $J_1-J_2$. As already known, we observe that the nearest neighbor concurrence of the ground state ($p=0$) does not show any discontinuity but the same for the first excited state ($p=1$) shows a sharp discontinuity in concurrence when plotted 
against $\alpha$. The same exercise for intermediate values of $p$
(say for $p=0.05$, $p=0.15$ and $p=0.25$) reveals an interesting observation  even if the probability of the presence of the first excited state is relatively small, the jump in the concurrence is 
discernible. So concurrence can be an indicator of phase transition even when the system resides in the first excited state with only a small probability and in the ground state with a large probability. See Figs. \ref{fig:N20_1}-(a) and \ref{fig:N20_1}-(b).
\begin{figure*}
\centering
\includegraphics[width=8.7cm]{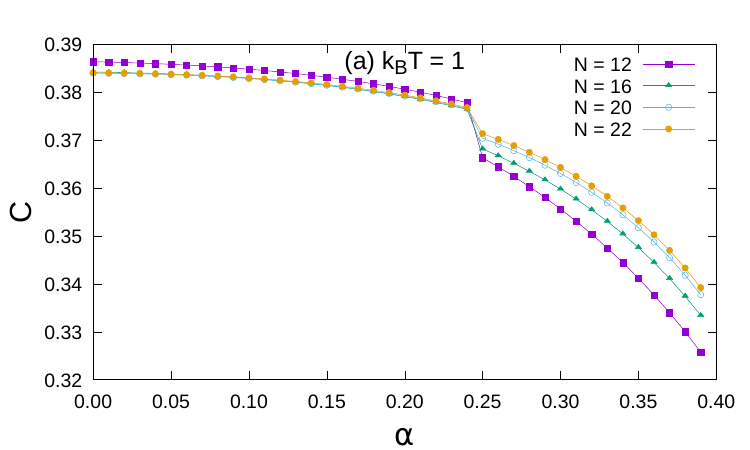}
\includegraphics[width=8.7cm]{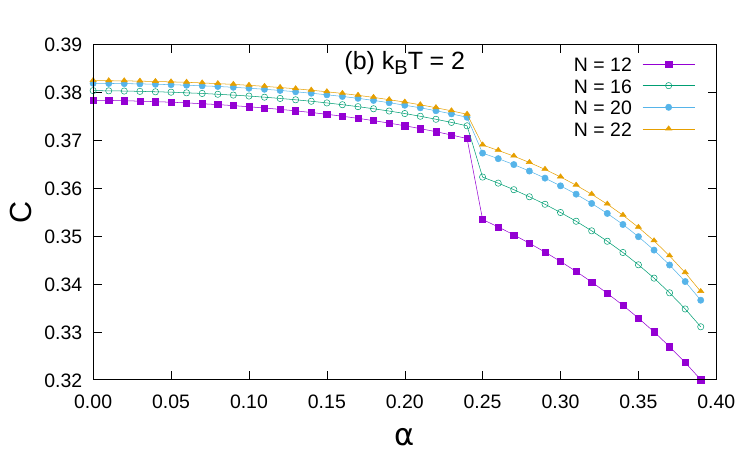}
\caption{\textcolor{black}{Nearest-neighbour concurrence of $J_1-J_2$ model for different system sizes at constant temperature. The connotations here are the same as in Fig.~\ref{fig:1p20}, but at a constant temperature with a varying mixing probability at each $\alpha$ for different system sizes. 
Here, in panel (a) $k_BT = 1$, and in panel (b) $k_BT = 2$ (in units of $J_1 \hbar$). The quantity plotted along the horizontal axes are dimensionless and the same along the vertical axes are in ebits.}}
\label{fig:aVc}
\end{figure*}
With increase in $p$, the jump increases and becomes more prominent. The point of discontinuity, which we consider to be the point of phase transition ($\alpha_c$) is the same for all values of $p$ for a fixed $N$. The neighbourhood, of the phase transition point is magnified and plotted in the insets of both the panels. The nature of concurrence is highly oscillating in this near-critical regime and the amplitude of oscillation increases with the increase of $p$ values. For determining the critical point, $\alpha_c$, we take the average of the two $\alpha$ values just before and just after the discontinuous jump in the concurrence. See the insets of Fig. \ref{fig:N20_1}. 
\par

In Fig. \ref{fig:1p20} we consider 
the same 
phenomenon but from a different perspective. 
Here we fix the values of $p$ of the subjacent state in each panel and investigate the nature of concurrence for different values of $N$. This time 
the figure reveals a different characteristic of the transition point. For a fixed $N$, the critical transition point ($\alpha_c$) is independent of $p$ (see Fig.~\ref{fig:N20_1}), whereas $\alpha_c$ decreases as the system size $N$ increases, for a fixed value of $p$. A tabular representation as a proof of this fact is presented in Table \ref{table:1}.
\par

\textcolor{black}{Note that the discrete jump in nearest-neighbor concurrence only occurs when applying periodic boundary conditions. When we apply an open boundary condition to the same model described in Eq. (\ref{eq:H}), specifically by setting the range of the first summation in Eq. (\ref{eq:H}) to run from $i=1$ to $N-1$ and the range of the second summation to run from $i=1$ to $N-2$, we observe that the nearest-neighbour concurrence of the subjacent state monotonically increases for system sizes that are twice odd numbers (e.g., $N = 10, 14, 18$), while it monotonically decreases for system sizes that are twice even numbers (e.g., $N = 12, 16, 20$). We do not detect any discontinuity in the concurrence within the region of $J_2/J_1 = \alpha \in [0.0, 0.5]$ for these cases. This may be because of the finite size effects of the quantum chain with open boundaries.}

\begin{figure}
    \centering
    \includegraphics[width=8.7cm]{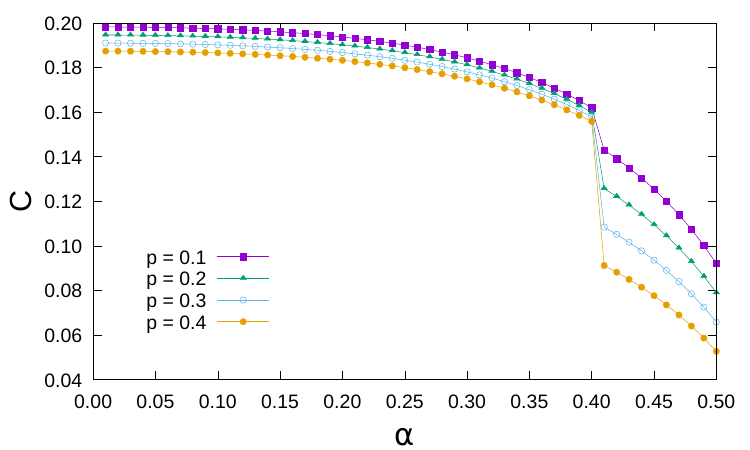}
    \caption{\textcolor{black}{Nearest-neighbour concurrence of the two-dimensional $J_1-J_2$ model for $N=16$ on a $4 \times 4$ lattice. Here we have depicted the nearest-neighbour concurrence with relative coupling strength $\alpha$, for different values of the mixing probability $p$ of the subjacent state. The quantity plotted along the horizontal axis is dimensionless and the same along the vertical axis is in ebits. }}
    \label{fig:2D_J1J2}
\end{figure}

\begin{table}[h!]
\begin{center}
 \begin{tabular}{||c c | c c||} 
 \hline
\hspace*{4mm} $N$ \hspace*{4mm}& $\alpha_c$ \hspace*{4mm}&\hspace*{4mm} $N$  &\hspace*{4mm} $\alpha_c$ \hspace*{4mm} \\ [0.5ex] 
 \hline\hline
 \hspace*{4mm}8 \hspace*{4mm}& 0.24630 \hspace*{4mm} &\hspace*{4mm}18  &\hspace*{4mm} 0.24221 \hspace*{4mm}\\ 
 \hspace*{4mm}10 \hspace*{4mm}& 0.24449 \hspace*{4mm}&\hspace*{4mm} 20 &\hspace*{4mm} 0.24201 \hspace*{4mm}\\
 \hspace*{4mm}12 \hspace*{4mm}& 0.24349 \hspace*{4mm}&\hspace*{4mm} 22 &\hspace*{4mm} 0.24180 \hspace*{4mm}\\
\hspace*{4mm} 14 \hspace*{4mm}& 0.24286 \hspace*{4mm}&\hspace*{4mm} 24 &\hspace*{4mm} 0.24164 \hspace*{4mm}\\
 \hspace*{4mm} 16 \hspace*{4mm}& 0.24248 \hspace*{4mm}&\hspace*{4mm}  &\hspace*{4mm}   \\[1ex] 
 \hline \hline 
\end{tabular}
\end{center}  
\caption{The values of $\alpha_c$ for increasing size $(N)$ of the system  for $p=0.30$ in the subjacent state for the $J_1-J_2$ model. All quantities are dimensionless.}
\label{table:1}
\end{table}
\textcolor{black}{This approach of utilising the subjacent state to detect the phase transition in the \(J_1-J_2\) Heisenberg spin chain can be extended to the $J_1-J_2$ model on a two-dimensional lattice. The two-dimensional $J_1-J_2$ Heisenberg spin model is governed by the Hamiltonian,
\begin{equation}
    H_{2-d} = J_1 \hbar\sum \vec{\sigma}_i \cdot \vec{\sigma}_j + J_2 \hbar \sum \vec{\sigma}_i \cdot \vec{\sigma}_k.
\end{equation}
Here the first summation involves pairs of nearest neighbours, whereas the second summation deals with pairs of nearest neighbours along the diagonals. In Fig.~\ref{fig:2D_J1J2}, we plot the nearest-neighbour concurrence of this two-dimensional $J_1-J_2$ model, for $N=16$, on a $4 \times 4$ lattice. Our findings reveal a discontinuous jump in the nearest-neighbour concurrence at approximately $\alpha \approx 0.4$, which indicates a phase transition from `ordinary-N\'{e}el order' to an intermediate phase of a `plaquette or columnar dimer' phase. 
It can be shown that the same methodology also detects  the (anti-)ferromagnetic to paramagnetic phase transition in the one-dimensional transverse Ising model, 
where the concurrence attains a maximum at the critical point.}

Till now, we were considering the situation where the mixing probability $p$ is left constant while nearest neighbour concurrence is investigated with varying relative coupling strengths, $\alpha$. See Figs. ~\ref{fig:N20_1} and ~\ref{fig:1p20}. This implies that we require the temperature to be changed at each $\alpha$ to keep the mixing probability fixed with the changes of $\alpha$. 
Arguably, a more practical study is to investigate the nature of nearest neighbour concurrence at a fixed temperature. Therein, the probability $p$, of the subjacent state, change with $\alpha$, as the change in $\alpha$ leads to changes in $E^{(1)}(N,\alpha)$ and $E^{(0)}(N,\alpha)$. For our analysis, we fixed the temperature to $k_BT = 1 \text{ and } 2$. 
\textcolor{black}{At very low temperatures, the system's state would consist, for practical purposes, of a mixture of the ground state and the first-excited state. Essentially, the excited state serves as a source of thermal noise in the system. To model a situation where this thermal noise arises due to a non-zero temperature, we can envision the mixed state as being associated with a fixed temperature, or with a fixed admixture of the excited state. The former is considered in Figs.~\ref{fig:N20_1} and~\ref{fig:1p20}, while the latter is taken up in Fig.~\ref{fig:aVc}. We find that both these noise models lead to the same phase transition points for the parent Hamiltonian.}


\textcolor{black}{From Fig.~\ref{fig:1p20}, it is evident that by keeping the mixing probability ($p$) constant, we can observe the phase transition in the one-dimensional $J_1-J_2$ model as we vary $\alpha$. Since fixing the probability leads to temperature variations with changes in $\alpha$, this phase transition might be construed as being induced by temperature. However, as illustrated in Figs.~\ref{fig:1p20} and~\ref{fig:aVc}, the nearest-neighbor concurrence at fixed temperature and at fixed $p$ is similar in nature, and the estimated critical point of phase transition is also at the same point in the parameter space in both the cases. In the case of phase transitions induced by thermal fluctuations, the phase transition becomes apparent as the system's temperature varies. Given that the critical point remains constant in scenarios where the temperature is fixed, it follows that the observed phase transitions are not induced by thermal fluctuations. Therefore, we can infer that the phase transition we observe here is not thermally induced.
}

Further illustrations are done for fixed $p$. Investigations at fixed temperature will provide equivalent information.
The difference between the values of concurrence $\Delta C$, just before and after the transition point, decreases as we increase the system size. The numerical values of $\Delta C$s for increasing system size, for a fixed $p$, are given in Table \ref{table:2}. This may potentially be a disadvantage 
in the thermodynamic limit, $N \rightarrow \infty$. 
A finite-size scaling is needed to see whether concurrence can still play the role of a good detector of phase transition in a mixture of ground and first excited state for $N \rightarrow \infty$. \par
\begin{table}[!htb]
\begin{center}
 \begin{tabular}{||c c | c c||} 
 \hline 
\hspace*{4mm} $N$ \hspace*{4mm}& $\Delta C$ \hspace*{4mm}&\hspace*{4mm} $N$  &\hspace*{4mm} $\Delta C$ \hspace*{4mm} \\ [0.5ex] 
 \hline\hline
 \hspace*{4mm}8 \hspace*{4mm}& $3.2 \times 10^{-2}$ \hspace*{4mm} &\hspace*{4mm}18  &\hspace*{4mm} $6.8\times 10^{-3}$ \hspace*{4mm}\\ 
 \hspace*{4mm}10 \hspace*{4mm}& $2.0\times 10^{-2}$ \hspace*{4mm}&\hspace*{4mm} 20 &\hspace*{4mm} $5.6\times 10^{-3}$ \hspace*{4mm}\\
 \hspace*{4mm}12 \hspace*{4mm}& $1.4\times 10^{-2}$ \hspace*{4mm}&\hspace*{4mm} 22 &\hspace*{4mm} $4.8\times 10^{-3}$ \hspace*{4mm}\\
\hspace*{4mm} 14 \hspace*{4mm}& $1.1\times 10^{-2}$ \hspace*{4mm}&\hspace*{4mm} 24 &\hspace*{4mm} $4.2\times 10^{-3}$ \hspace*{4mm}\\
 \hspace*{4mm} 16 \hspace*{4mm}& $8.3\times 10^{-3}$ \hspace*{4mm}&\hspace*{4mm}  &\hspace*{4mm}   \\[1ex] 
 \hline \hline 
\end{tabular}
\end{center}  
\caption{The values of $\Delta C$ (in ebits) for different sizes ($N$) of the system for $p = 0.30$, for the $J_1-J_2$ model.}
\label{table:2}
\end{table} 
\begin{figure}[!htb] 
\includegraphics[width=8.7cm]{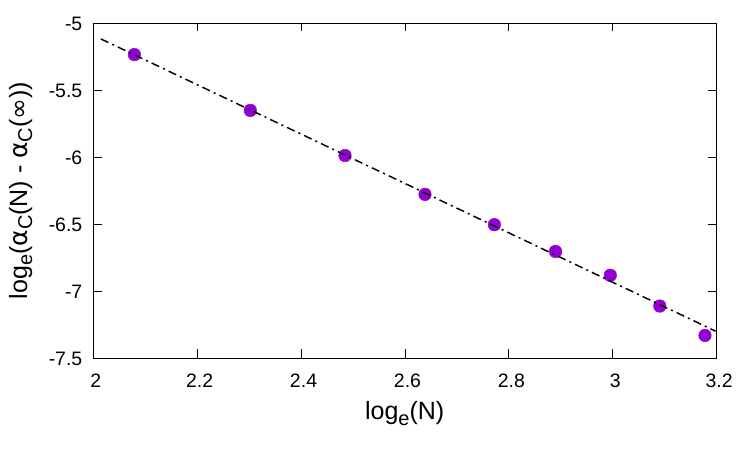}
\caption{Critical point of the $J_1-J_2$ model in the thermodynamic limit for the subjacent state at $p = 0.30$. Here we have fitted the values of $\alpha_c$ corresponding to different $N$ obtained from Table~\ref{table:1}, with Eq.~\eqref{eq:fit1}. The data points are shown by blue dots and the black dotted line represents the fitting curve. Both the axes are plotted in $\log$ scale and the quantities plotted are dimensionless.}
\label{fig:alphaC}
\end{figure}
\begin{figure*} 
\centering
\includegraphics[width=8cm]{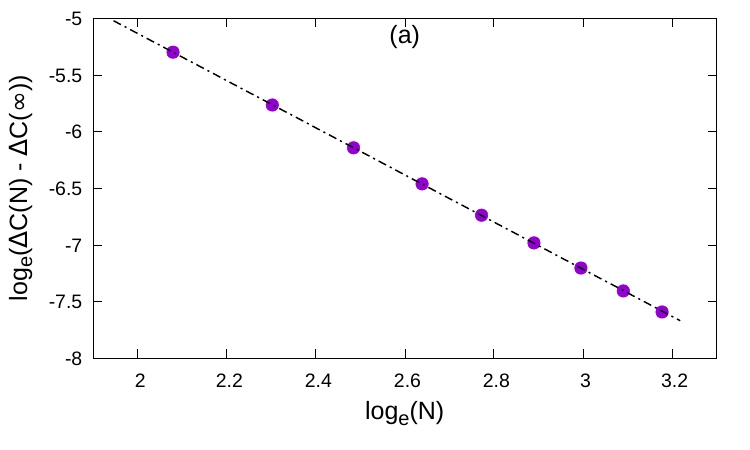}
\includegraphics[width=8cm]{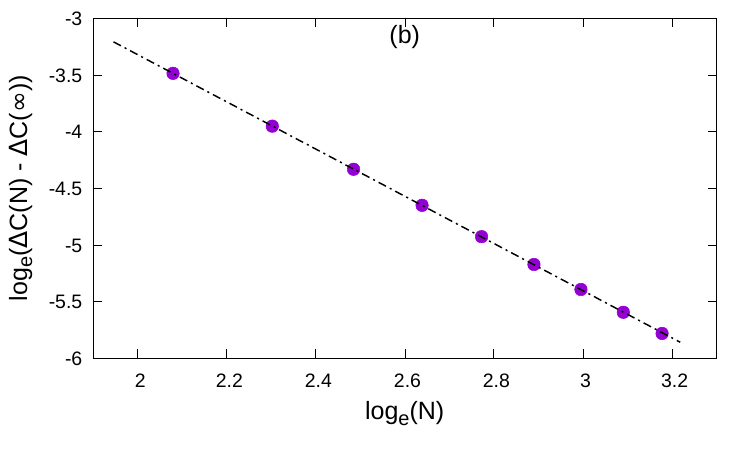}
\caption{Jump in concurrence at critical point at thermodynamic limit. Here we have fitted the values of $\Delta C$ for different $N$ with Eq.~\eqref{eq:fit2} for (a) $p=0.05$ and (b) $p=0.30$. The data points are shown by blue dots and the black dotted line represents the fitting curve. The axes are plotted in $\log$ scale and the quantity described along vertical axes are in ebits whereas the same along horizontal axes is dimensionless.}
\label{fig:DeltaC}
\end{figure*}

As the precise quantum phase transition point on the $\alpha$-axis for an infinitely large system is unknown, we extrapolate the transition points of $N=8,10,12,14,16,18,20,22,24$ size systems to $N=\infty$. To extrapolate, we have chosen the fitting function, viz. 
\begin{equation} \label{eq:fit1}
\alpha_c(N) = \alpha_c(\infty) + a N^{-b}.
\end{equation}
On fitting the values of $\alpha_c(N)$ with the system size ($N$) from Table \ref{table:1}, we obtain $\alpha_c(\infty) = 0.24099$ with the 95\% confidence interval $\pm 4.9 
\times 10^{-9}$, $a = 0.24$ 
(95\% confidence interval $\pm 1.9 
\times 10^{-6}$) and $b = 1.8$ 
(95\% confidence interval $\pm 4.1 
\times 10^{-6}$). Thus the value of relative coupling ($\alpha$) at which there is a discontinuity, asymptotically tends to $\alpha_c = 0.24099$, with the standard error 
$ 3.2 
\times 10^{-5}$, for $p = 0.30$ 
in the subjacent state. Here we have used non-linear least-square fitting to find the values of the parameters, 
the $95\%$ confidence intervals, and the error, for the fitting curves~\cite{Bates1988,Press2007,Sehrawat2021}. We can also see from Fig. \ref{fig:alphaC}  that the difference, $\alpha_c(N)-\alpha_c(\infty)$, decreases almost linearly with the increase of $N$, when plotted in $\log-\log$ scale.

{Having obtained the transition point (for $p = 0.30$) in the thermodynamic limit, we can now investigate} the utility of concurrence as a phase transition detector at $N \rightarrow \infty$.
For checking this, 
we fit the values of $\Delta C$ for different $p$'s with the following expression:
\begin{equation} \label{eq:fit2}
\Delta C(N) = \Delta C(\infty) + \gamma N ^{-\delta}.
\end{equation} 
We find that for $p = 0.30$ as well as for other values of $p$, the jump does not asymptotically tend to zero. There is some finite discontinuity $\Delta C(\infty)$ as $N \rightarrow \infty$ and hence we have achieved our goal of finding a quantum information-theoretic indicator of phase transition in the thermodynamic limit. The numerical values of the jumps in the thermodynamic limit, $\Delta C(\infty)$, for different values of mixing probability, along with the standard errors of fitting, are given in  Table~\ref{table:3}. \textcolor{black}{The scaling exponent $\delta$ in \eqref{eq:fit2}, turns out to be approximately 2.08 for all values of $p$}.  
\begin{table}[h!]
\begin{center}
 \begin{tabular}{||c c c||} 
 \hline 
\hspace*{4mm} $p$ \hspace*{4mm}& $\Delta C(\infty)$ \hspace*{4mm}& Error \hspace{4mm}\\ [0.5ex] 
 \hline\hline
\hspace*{4mm}0.01 \hspace*{4mm}& $8.3\times 10^{-4}$ \hspace*{4mm} & $2.9\times 10^{-7}$\hspace*{4mm}\\ 
\hspace*{4mm}0.05 \hspace*{4mm}& $8.6\times 10^{-4}$ \hspace*{4mm} & $4.1\times 10^{-6}$\hspace*{4mm}\\
\hspace*{4mm}0.10 \hspace*{4mm}& $9.1\times 10^{  -4}$ \hspace*{4mm} & $9.1\times 10^{-6}$\hspace*{4mm}\\
\hspace*{4mm}0.15 \hspace*{4mm}& $9.6\times 10^{-4}$ \hspace*{4mm} & $1.5\times 10^{-5}$\hspace*{4mm}\\
\hspace*{4mm}0.20 \hspace*{4mm}& $1.0\times 10^{-3}$ \hspace*{4mm} & $1.7\times 10^{-5}$\hspace*{4mm}\\
\hspace*{4mm}0.25 \hspace*{4mm}& $1.1\times 10^{-3}$ \hspace*{4mm} & $2.5\times 10^{-5}$\hspace*{4mm}\\
\hspace*{4mm}0.30 \hspace*{4mm}& $1.1\times 10^{-3}$ \hspace*{4mm} & $1.8\times 10^{-5}$\hspace*{4mm}\\
\hspace*{4mm}1.00 \hspace*{4mm}& $1.8\times 10^{-3}$ \hspace*{4mm} & $8.4\times 10^{-5}$\hspace*{4mm}\\

\hline \hline 
\end{tabular}
\end{center}
\caption{The asymptotic values of $\Delta C$ (in ebits) in the thermodynamic limit ($N \rightarrow \infty$) for different values of $p$, for the $J_1-J_2$ model. }
\label{table:3}
\end{table}
The fitting functions and the corresponding $\Delta C(N)-\Delta C(\infty)$ (in $\log-\log$ scale) with respect to system size $N$ for two fixed values of $p$ is demonstrated in Fig. \ref{fig:DeltaC}. {We have performed the analysis also for other values of $p$ and found a qualitatively similar behaviour.} The behaviour is very similar to that of $\alpha_c(N)-\alpha_c(\infty)$. Both decreases almost linearly with the increase of $N$ on the $\log-\log$ scale. Compare Figs.~\ref{fig:alphaC} and~\ref{fig:DeltaC}.

\begin{figure} 
\centering
\includegraphics[width=8cm]{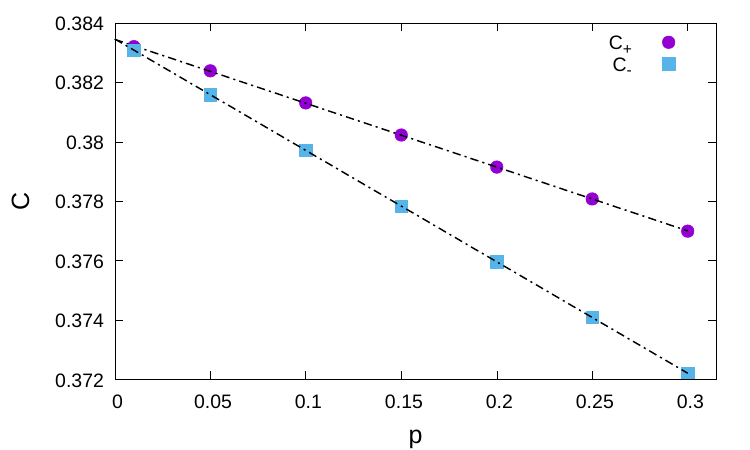}
\caption{Nearest neighbour concurrence just before and just after the jump near the critical point of the $J_1-J_2$ model for the subjacent state. Here we have looked into the dependence of $C_+$ and $C_-$ on $p$ for $N=20$. The rectangular dots are the values before the discontinuity ($C_-$) and the circular dots are the same just after the jump ($C_+$). The dotted lines represent the corresponding fitting functions. The quantity plotted along the horizontal axis is dimensionless and the same described along the vertical axis is in ebits.}
\label{fig:Cptr}
\end{figure}

Till now we were discussing about the phase transition point of the $J_1-J_2$ model, and the jump in concurrence at that point.
It could also be interesting to study the values of nearest neighbour concurrence in the vicinity of phase transition point and its dependence on the mixing probability $p$ for different system sizes. 
We depict the dependence of concurrence, just before the jump by $C_-$ and the same just after the jump by $C_+$, on $p$ for $N = 20$ in Fig.~\ref{fig:Cptr}, with rectangular and circular dots respectively. We find that for small values of $p$, the values of concurrence, either way about the point of discontinuity, decreases linearly. {The behaviour is similar for other values of $N$ too.}
Here we end our discussion about the phase transition of the ideal $J_1-J_2$ Heisenberg spin chain. In the succeeding sections, we will look into the response to anisotropy and disorder of the phase diagram of the model.
\section{Response to anisotropy of the phase transition point}
\label{Sec:4}
\begin{figure*}
\includegraphics[width=7.5cm]{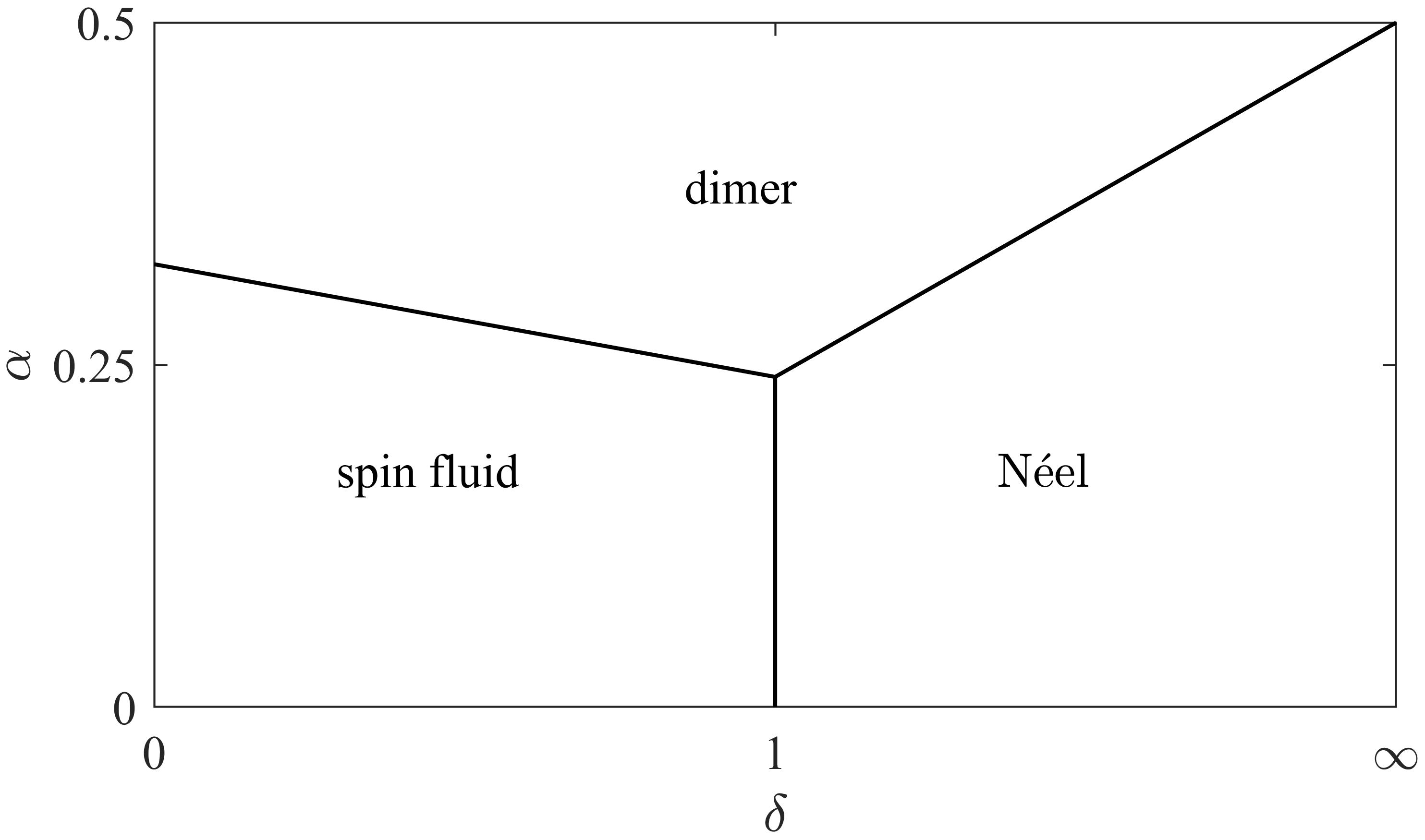}
\includegraphics[width=9.5cm]{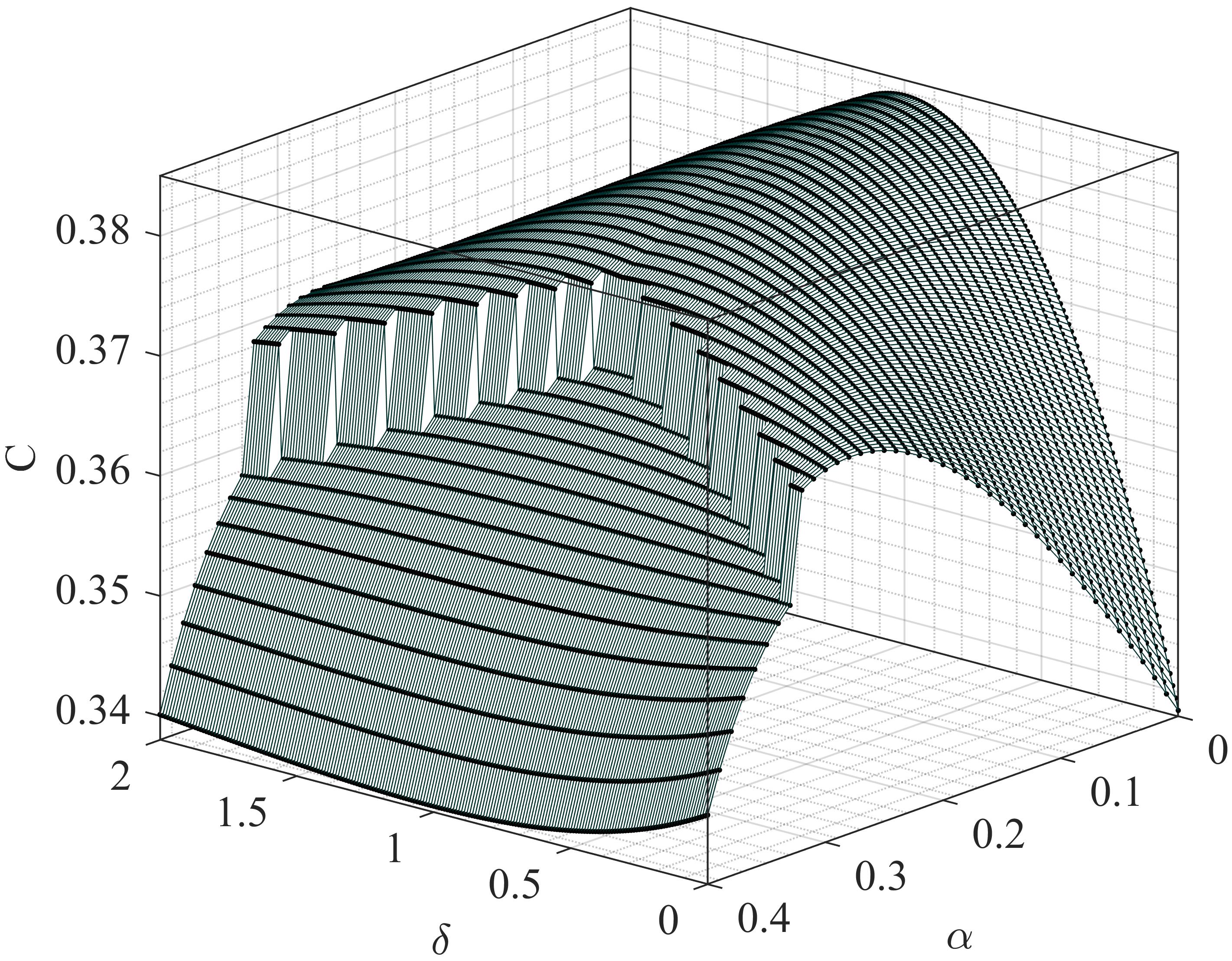}
\caption{Phase diagram of the XXZ $J_1-J_2$ spin model. The left panel represents a schematic quantum phase diagram of the one-dimensional XXZ $J_1-J_2$ spin model on the $(\alpha$ , $\delta)$ plane. In the right panel, we have plotted the nearest neighbour concurrence of the subjacent state on the $(\alpha,\delta)$ plane, and considered it as an order parameter {for the transitions in the XXZ $J_1-J_2$ model. Here, $N = 16$.} The concurrence $(C)$ is in ebits, while all other quantities are dimensionless.}
\label{phase_diagram_xxz}
\end{figure*}

The one-dimensional $J_1-J_2$ Heisenberg spin model with an anisotropic coupling constant along $z$-direction can be named as the
one-dimensional spin-$\frac{1}{2}$ $XXZ$ $J_1-J_2$ spin model 
and its governing Hamiltonian may be written as
\begin{equation}
H_{XXZ} =J_1 \hbar \sum_{i=1}^{N}(\vec{\sigma}_i \cdot \vec{\sigma}_{i+1})_\delta+J_2 \hbar \sum_{i=1}^{N}(\vec{\sigma}_i \cdot \vec{\sigma}_{i+2})_\delta,
\label{hamiltonian_1d}
\end{equation}
where $(\overrightarrow{\sigma_i}\cdot\overrightarrow{\sigma_j})_\delta=\sigma_i^x \sigma_j^x+\sigma_i^y \sigma_j^y+\delta \sigma_i^z \sigma_j^z$ with $\delta$ being a dimensionless anisotropy constant. 
This model is known to have various quantum phases~\cite{10.1143/PTPS.145.113} and can describe the spin-Peierls compound CuGeO$_3$~\cite{Hase1993,Castilla1995}. As already discussed, the system is in a spin fluid state for $\alpha \lesssim 0.24$ and $\delta < 1$, and it goes to  a N\'{e}el phase for $\delta >1$. For $\alpha \gtrsim 0.24$, the system goes into a dimer phase. A schematic representation of these phase boundaries is shown in the left panel of Fig.~\ref{phase_diagram_xxz}.

\begin{figure*}
\includegraphics[width=8 cm]{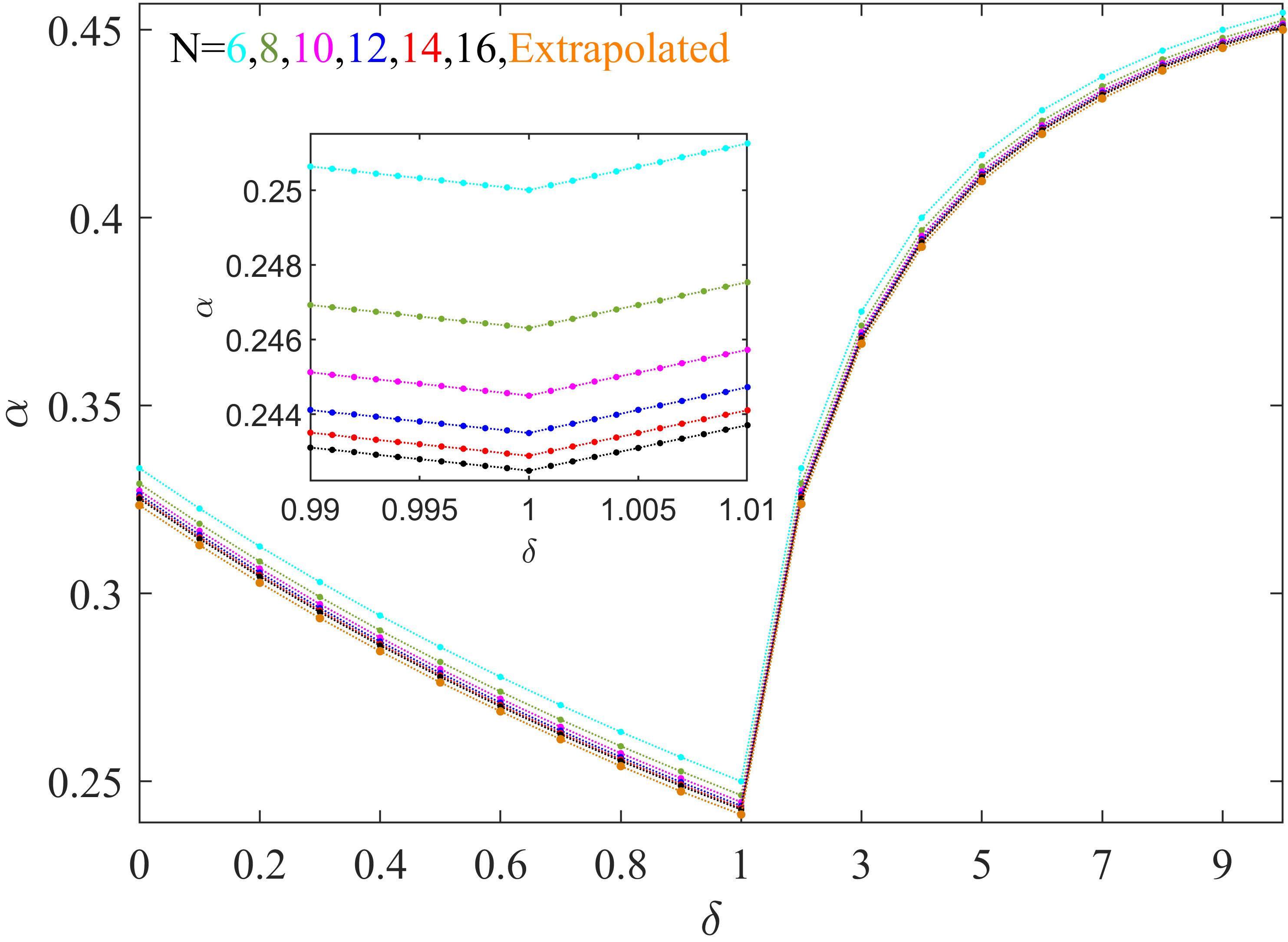}
\includegraphics[width=8 cm]{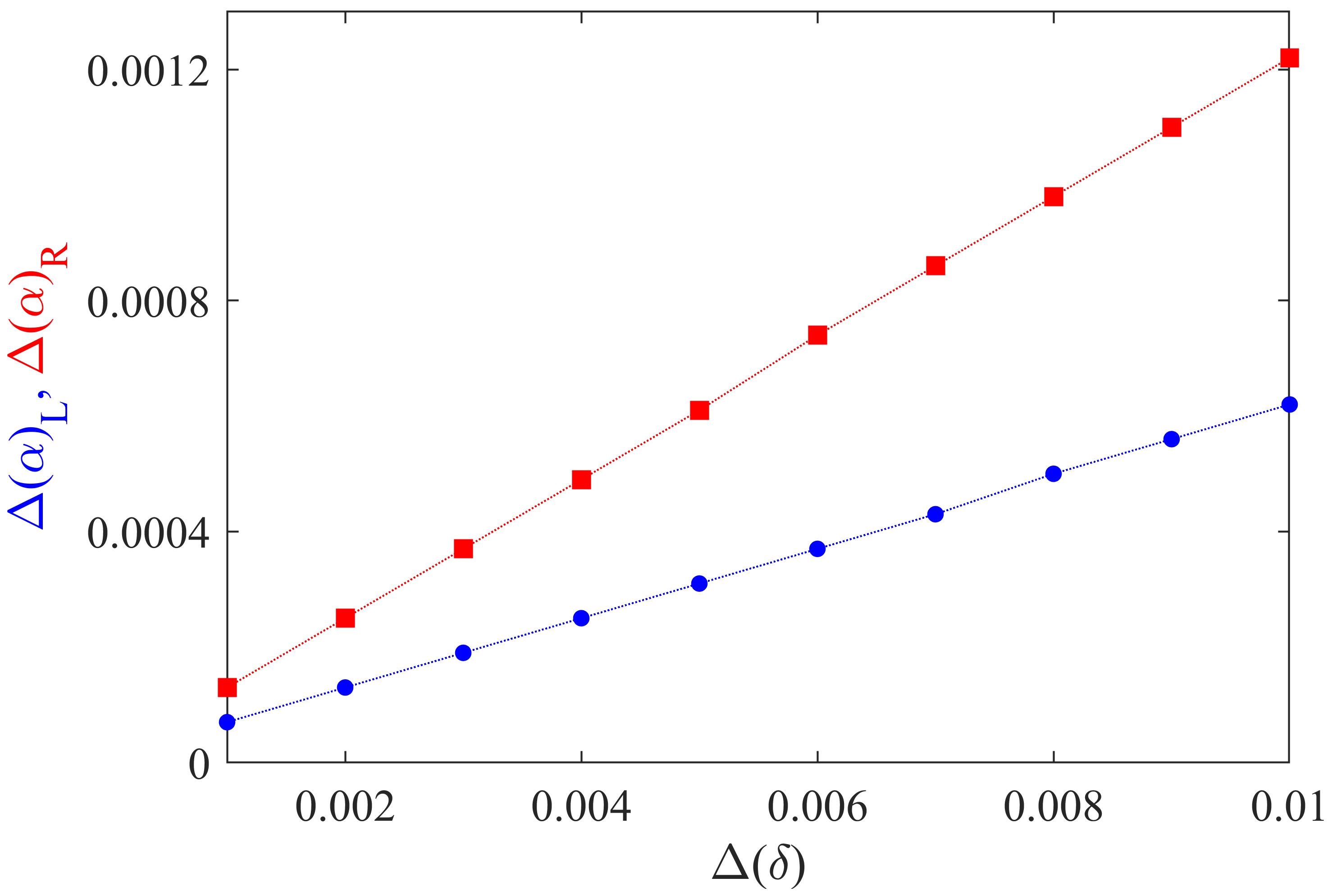}
\includegraphics[width=8 cm]{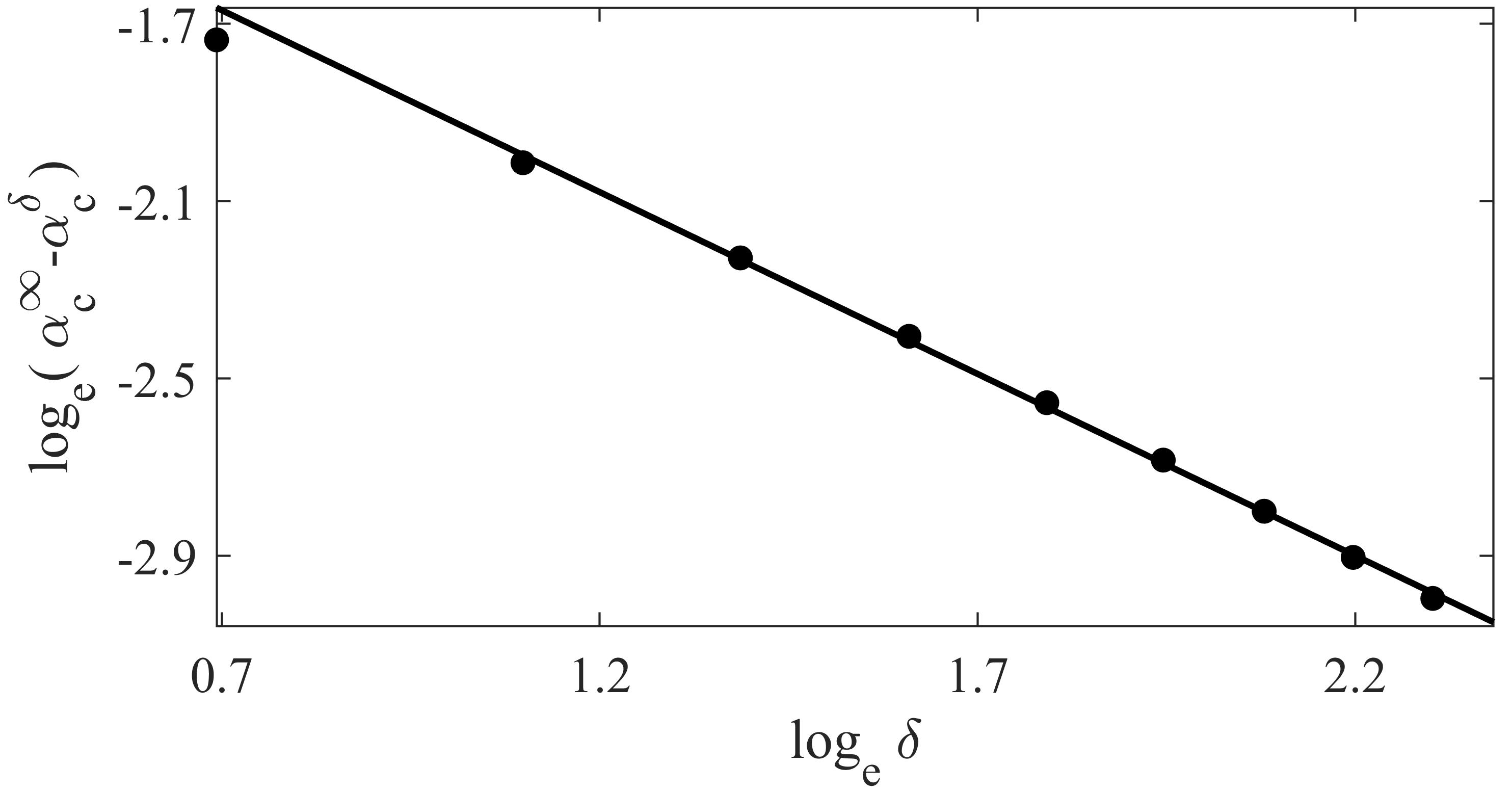}
\caption{{Phase diagram and finite-size scaling for transitions of the XXZ $J_1-J_2$ model. The phase diagram is presented in the upper left panel. We focus on the transitions to the dimer phase and the multi-critical point. The transitions are assumed to be signaled by the discontinuities of the nearest neighbour concurrence in the subjacent state.}
The orange dots correspond to the extrapolated discontinuity points in the large $N$ limit. A magnified version of this depiction closer to the $\delta=1$ region is shown in the inset. 
In the upper right panel, we compare the nature of critical transition points on the two sides of $\delta=1$. Here we take the magnitude of the distance from $\delta=1$ as $\Delta(\delta)$ and depict the corresponding $\alpha_c$ values for both sides of the $\delta=1$ point. The critical $\alpha_c$ values on the left of $\delta=1$ is depicted by blue dots and the same on the right side are depicted by red squares for system size $N=16$. 
In the lower panel, we present the finite size scaling of the extrapolated critical transition points $\alpha_c$ along $\delta$-axis for $\delta>1$ for $N \rightarrow \infty$.  
For the values of scaling exponent and the corresponding $95\%$ confidence interval see the main text. All the quantities plotted along the axes in all the three panels are dimensionless. 
}
\label{con_scaling}
\end{figure*}

In the previous section, we have studied the behaviour of concurrence by changing the mixing probability. In this section, we have taken a fixed mixing probability $p$. All further analysis is done for 
\begin{equation}
    p = 0.2689 \approx \frac{e^{-1}}{1 + e^{-1}}.
    \label{Eq.(HRI)}
\end{equation}
We are therefore fixing the temperature as $T \approx \frac{E^{(1)}(N,\alpha)-E^{(0)}(N,\alpha)}{k_B}$ for each $\alpha$. Now we investigate the behaviour of concurrence on the $(\alpha, \delta)$ parameter space. In the right panel of Fig.~\ref{phase_diagram_xxz}, we observe that there is a dip in the concurrence along the $\delta=1$ line for low $\alpha$, indicating the spin fluid to N\'{e}el order quantum phase transition. It shows a sharp drop to indicate the spin fluid to dimer quantum phase transition for $\delta<1$, and a sharp jump to indicate the dimer to N\'{e}el quantum phase transition for $\delta>1$. \textcolor{black}{Thus, we observe that a discontinuity in the concurrence serves as an indicator of a phase transition between the spin fluid and dimer phases in the anisotropic one-dimensional $J_1-J_2$ model. Also, the phase transition between N\'{e}el and dimer is also signified by discontinuity of concurrence. On the other hand, we observe that the phase transition between the spin-fluid and N\'{e}el phases is distinguished by a discontinuity in the first derivative of the concurrence. So, here the nature of the signatures observed in concurrence varies for different phase transitions. In this regard, it is important to note that, as the discontinuity in the order parameter is indicative of a phase transition, it is also quite common for order parameters to display continuity near the critical point, while their derivatives exhibit discontinuities, as observed in \cite{Syljuaasen2003, Wu2004}. 
}
{The points of transition, say along lines parallel to the $\alpha$-axis, are calculated by using exact diagonalisation on finite systems. A finite-size scaling analysis is therefore useful. For a fixed $\delta$, we fixed the critical point $\alpha_C(\delta, N)$, for a system of size $N$, and use the }
rational polynomial,
\begin{equation} \label{rational}
    f\left(N\right)=\frac{p_1N^2+p_2N+p_3}{N^2+q_1N+q_2},
\end{equation}
to estimate the quantum phase transition point of an infinitely large system ~\cite{NISTESH}. 
The corresponding phase diagram, using both finite-size predictions as well as the extrapolated ones is presented in the upper left panel of Fig.~\ref{con_scaling}. 
We analyse their finite-size scaling behavior, 
by fitting a function of the form,
\begin{equation} 
\label{exponent}
    	\textcolor{black}{\alpha_c(N)=\alpha_c(\infty)+a'N^{-\beta}},
\end{equation}
and obtain the 
finite size scaling exponents $(\beta)$ for different $(\delta)$. 
The corresponding data for $\delta < 1$ is presented in ~\ref{delta0to1} and the same for $\delta \ge 1$ is shown in ~\ref{delta1to10}, {in the Appendix \ref{appendix1}}. Here $\alpha_c(\infty)$ is the value of $\alpha_c$ for $N \rightarrow \infty$ and $a'$ is a constant. 
We magnify the region around the multi-criticality point for $\alpha\approx 0.24$ and $\delta=1$, in the inset of {of the upper left panel of Fig.~\ref{con_scaling}}. 
\begin{figure*}[!htb] 
\centering
\includegraphics[width=8cm]{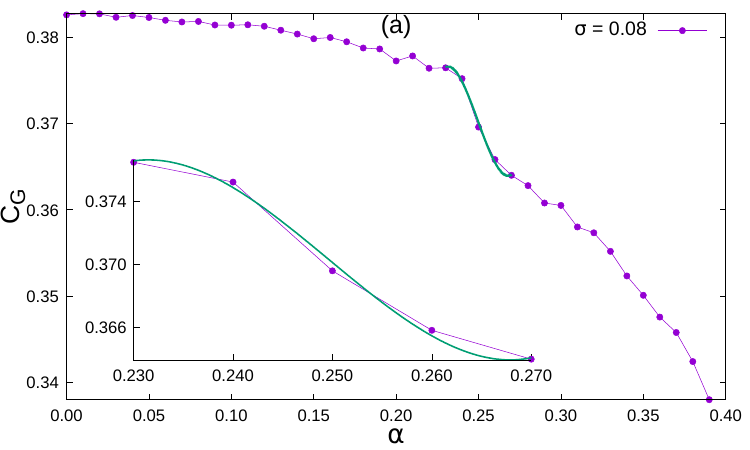}
\includegraphics[width=8cm]{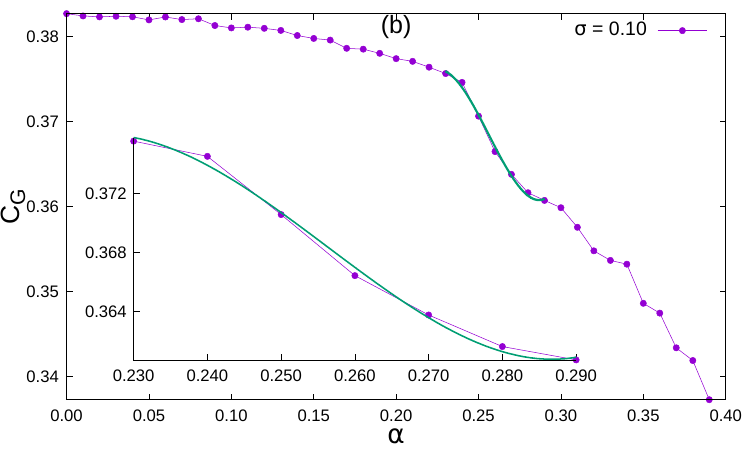}
\includegraphics[width=8cm]{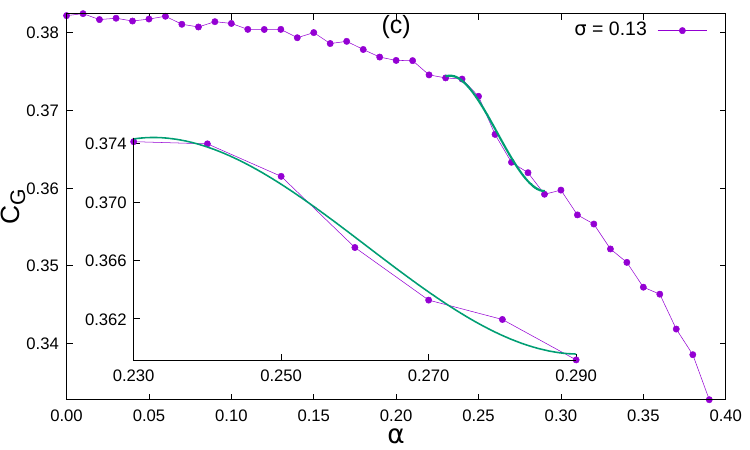}
\includegraphics[width=8cm]{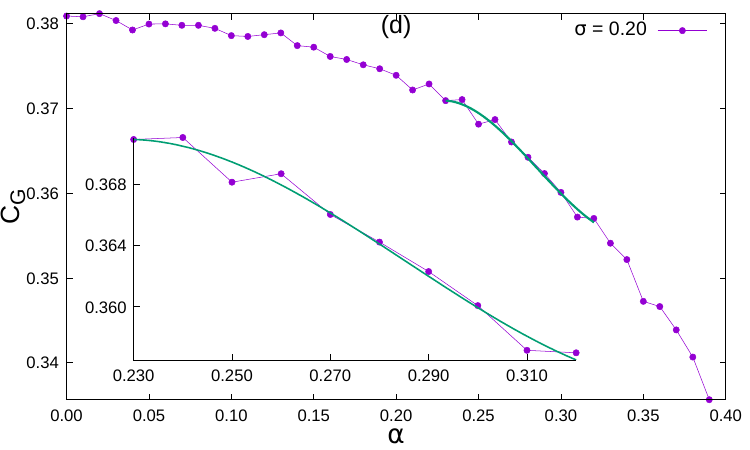}
\caption{Nearest neighbour concurrence in presence of Gaussian disorder for $N=14$. The disorder averaged concurrence (averaged over $5\times 10^4$ realizations) is plotted along the vertical axes when the $\Delta$s are chosen independently from a Gaussian distribution with mean zero and for different values of the standard deviation $\sigma$ (given in the legends). The region close to the transition point is shown in the insets where the continuous line represents the polynomial fitting function. The quantity plotted along horizontal axes is dimensionless, whereas that plotted along vertical axes is in ebits.}
\label{fig:14dis}
\end{figure*}
We notice that the phase boundary changes with the change in $\delta$. The numerical values of critical phase transition points $\alpha_c$ decrease monotonically with the increase of $\delta$ from 0 to 1 and increase monotonically thereafter. 

For further analysis, we plot the critical transition points $\alpha_c$ with respect to the distance in $\delta$ axis from the point $\delta=1$ in the upper right panel of Fig.~\ref{con_scaling}, for $N=16$. From the inset of the upper left panel of Fig.~\ref{con_scaling}, we can observe an asymmetry in $\alpha_c$ values on the two sides of $\delta=1$. This asymmetry is clearer in the upper right panel. 
We can see that the slope of the curves representing the $\Delta\alpha$ values for the $\delta > 1$ region is much greater than those for the $\delta < 1$ one. 

We find that $\alpha_c$ approaches $\frac{1}{2}$ in the limit of $\delta \rightarrow \infty$.
We now analyse the scaling behavior of the phase transition point with $\delta$. 
We choose the fitting function,
\begin{equation} \label{exponent_delta}
    	\alpha_c^{\delta}=\alpha_c^{\infty}-a''\delta^{-d},
\end{equation}
where $\alpha_c^{\infty}$ is the value of $\alpha_c$ at $\delta=\infty$ and $a''$ is a constant. 
We present the variation of $\alpha_c^{\infty}-\alpha_c^{\delta}$ against $\delta$ (both on $\log$ scale) for $N \rightarrow \infty$ 
in the lower panel of Fig.~\ref{con_scaling}. 
The scaling exponent is found to be $d=0.82$ 
with
95\% confidence interval $\pm 0.025$ 
and the standard error is $0.012$ 

\section{Response to glassy disorder of the phase transition point}
\label{Sec:5}
A system is said to be glassy disordered if there is a disordered system parameter whose  equilibration time is much larger than the time spans relevant for our investigations of the system. So, during the time of observation, the disordered parameters effectively do not change for a certain realization of disorder. 
This disorder is analogous to that of a spin glass system~\cite{Chowdhury1986,Mezard1987,Nishimori2001}. Glassy disorder has also been referred to as a quenched disorder~\cite{sachdev_2011,Chowdhury1986,Mezard1987,Nishimori2001,Chakrabarti1996, suzuki2012quantum, Ferrari2022}. 
Studies on spin chains with glassy disorder include~\cite{Aharony1978,Lee1985,Binder1986,Niederberger2010,Kjall2014,Sadhukhan12016,PhysRevB.60.344,pub.1003333330,PhysRevLett.102.057205,Ghosh2020,Bera2016,Rakshit2017,Bera2019,Sarkar2022}.
Here we study the effect of introduction of a glassy disorder parameter in the $J_1-J_2$ model.
We analyse the response of the quantum phase transition point on introduction of a glassy anisotropy parameter. Precisely, we insert a $z-z$ interaction term in the $J_1-J_2$ Hamiltonian ~\eqref{eq:H} for each Heisenberg interaction term $\vec{\sigma}_i\cdot\vec{\sigma}_j$. The unit-free coupling strength, $\Delta$, for each interaction then is independently distributed with respect to that for any other term. However, they are all Gaussian distributed with vanishing mean and standard deviation $\sigma$:   
\begin{equation}
    P(\Delta)=\frac{1}{\sigma\sqrt{2\pi}}e^{-\frac{1}{2}\big(\frac{\Delta}{\sigma}\big)^2}, \quad -\infty<\Delta<\infty.
\end{equation}
{The considerations in this section are therefore complementary to those in the preceding one, even though both deal with a z-z interaction anisotropy in the $J_1-J_2$ model. Note that the usual $J_1-J_2$ model lies in the $\sigma \rightarrow 0$ limit for the analysis of the glassy disordered model of this section. In comparison, the same limit is obtained for $\delta \rightarrow 1$ for the considerations of the XXZ $J_1-J_2$ model of the preceding section.  
The disorder averaged concurrence is given by}
\begin{equation}
    C_G=\int_{-\infty}^{\infty} C(\tilde{\Delta}) \frac{1}{\sigma\sqrt{2\pi}}e^{-\frac{1}{2}\big(\frac{\tilde{\Delta}}{\sigma}\big)^2}d\tilde{\Delta}.
\end{equation}
{Here, $\tilde{\Delta}$ represents the entire collection of $\Delta$s of the system, and the $\tilde{\Delta}^2$ in the exponent denotes the sum of the square of such $\Delta$s.}
We use  Monte Carlo integration to evaluate the integral. 
Now for each realisation, $\Delta$s are chosen independently by choosing random numbers from a Gaussian distribution with mean zero and standard deviation $\sigma$, and {then the nearest neighbour concurrence is calculated for that realisation by using the subjacent state. We average over a large number of realisation and check for convergence.}

In presence of the glassy disorder in the parameter $\Delta$, the concurrence shows a behaviour similar to that of the ordered case and a sufficiently visible difference in the values of concurrence before and after the phase transition point is observed. See Fig.~\ref{fig:14dis} in this regard. The difference is that the change in concurrence around the phase transition point is not as sharp as that in the ideal situation. The discontinuity in concurrence at the phase transition point becomes less and less prominent, as we increase the strength of disorder, ie., the standard deviation $\sigma$ of the disorder distribution. As we increase $\sigma$ to more than $0.2$, it becomes difficult to point out the region around the phase transition point.

{Since the discontinuity in the profile of the concurrence is not sharp, we adopt here a method that is different from the one used until now to identify the phase transition.} As we see from Fig.
~\ref{fig:14dis}, the profile of concurrence 
with $\alpha$ goes from concave to convex at the phase transition point. 
We can then fit a cubic polynomial in the near-critical region, and the inflection point of the polynomial can be considered as the point of phase transition.
The inflection point of a cubic polynomial is the point where the function changes its curvature from negative to positive or vice versa, and hence the polynomial has a vanishing double derivative at this point. Suppose that the cubic polynomial is given by
\begin{equation}
\label{eq:fitting_poly}
C_G(\alpha)= a_1\alpha^3 + b_1\alpha^2 + c_1\alpha +d_1.
\end{equation}
The near-critical region along with the fitting function are shown in the insets of Fig.
~\ref{fig:14dis}. 
The values of the critical points for different system sizes by changing the strength of disorder are given in Table~\ref{table:disorder} for the mixing probability $p = 0.2689$ (see \eqref{Eq.(HRI)}), and in Table~\ref{table:disorderp13} for mixing probability $p = 0.1344$,  along with the standard errors of fitting.
The phase transition point gradually shifts to higher $\alpha$ with the increase of the strength of disorder ($\sigma$). \textcolor{black}{Furthermore, we analyse the finite-size scaling for the case of $p = 0.2689$ from the data set in Table~\ref{table:disorder}. The fitting function is of the form, 
$$\alpha_C^{\text{dis}}(N) = \alpha_C^{\text{dis}}(\infty) + aN^{-b}.$$ 
For $\sigma = 0.05$, $\alpha_C^{\text{dis}}(\infty) = 0.2476$ and $b = 7.69$ and for $\sigma = 0.08$, $\alpha_C^{\text{dis}}(\infty) = 0.2492$ and $b = 7.38$. For the former, the root-mean-square error of the fitting function is \(0.87 \times 10^{-3}\), and for the latter, the same is $0.94 \times 10^{-3}$.}

\textcolor{black}{Throughout this study, we have utilised the subjacent state to detect the phase transition of a $J_1-J_2$ Heisenberg spin chain. However, it is important to ask, if any mixed state other than the subjacent state, defined in Eq.~\eqref{eq:state}, can show signatures of phase transition. In the work presented in~\cite{Biswas2022}, it was demonstrated that, for the one-dimensional $J_1-J_2$ model, the thermal state, i.e., the canonical equilibrium state, which is a mixture of all possible energy eigenstates, exhibits no discontinuity in nearest neighbor concurrence near the critical point. Furthermore, the authors explored a scenario where the full multiplet was not included, and the mixture incorporated states up to the third excited state. In this particular case as well, they did not observe any discontinuity in the nearest neighbor concurrence near the critical point. Therefore, it becomes evident that employing the subjacent state is more advantageous in detecting quantum phase transitions in a $J_1-J_2$ Heisenberg spin chain compared to using the thermal state or states composed of higher energy levels.}




\section{Conclusion}
\label{Sec:6}
We have investigated the utility of nearest neighbour 
entanglement of a mixture of the ground and first excited states - the subjacent states - which mimic the low but non-zero temperature states of the system, for detection and analysis of quantum phase transitions in one-dimensional $J_1-J_2$ Heisenberg quantum spin models.
We want to find signatures of 
the ground state quantum phase transitions in a given physical system at low but finite temperatures by utilizing the subjacent states. This is of significant interest because a real physical system will always be at a finite temperature. We try to find the extent to which such a finite-temperature state can mirror the zero-temperature transitions in the system. 
We have studied the dependencies of the critical phase transition point on the mixing probability, and found that 
concurrence - a measure of two-qubit entanglement - can play the role of a good order parameter even if the ground state probability is large compared to that of the first excited state, whereas it is already known that the ground state concurrence itself cannot perform the role of a good indicator of phase transition in these models. We have extrapolated the results to the thermodynamic limit and obtained the phase transition point $\alpha_c \approx 0.24099$, for the isotropic $J_1-J_2$ Heisenberg quantum spin chain with $\alpha = J_2/J_1$, which is very close to the 
value already known in literature, obtained by various methods~\cite{OKAMOTO1992433,PhysRevB.54.R9612, Alet2010, Biswas_2020} and we have also shown that the critical point does not depend on this mixing probability.

Moreover, we applied the same order parameter of the subjacent state for analysing phase transitions in the $J_1-J_2$ model, tweaked by either anisotropy or glassy disorder in the coupling strengths.
We introduced anisotropy in one specific direction for both the nearest neighbour and next-nearest neighbour interactions, and studied the effect of anisotropy on the critical phase transition point for various system sizes, and subsequently extrapolated to the thermodynamic limit. Here, we obtained a phase diagram of the anisotropic $J_1-J_2$ model, which shows the phase boundaries of the existent phases of the system like spin fluid state, dimer phase and N\'{e}el phase. In previous works, these phase boundaries of an anisotropic $J_1-J_2$ model was detected by some other order parameters ~\cite{10.1143/PTPS.145.113,Sato2011COMPETINGPI}, but we showed that detection of such phase boundaries is possible at finite temperature with a quantum-informatics order parameter. 

Separately, we incorporated a glassy disorder parameter in a coupling of the $J_1-J_2$ model and investigated its effect on the phase diagram. We observed that the disorder  
smoothens the marker of phase transition.
The phase transition point also shifted towards higher values of $\alpha$ depending on the strength of the disorder. Importantly, we performed finite-size scaling in each of the cases considered. The phenomenon of phase transition of a disordered $J_1-J_2$ model is already studied in the  literature, but having the nearest neighbour entanglement as an indicator of phase transition, even in presence of disorder in this model, is a potentially interesting result, which reveals the usefulness of quantum information theoretic properties in detecting paradigmatic many-body phenomena in a realistic scenario. We believe that the results would lead towards the establishment of the  consideration of physical quantities in the subjacent state as a proper and fruitful order parameter for detection and characterisation of cooperative physical phenomena, since such quantities in the subjacent state are theoretically accessible just like ground state physical quantities and experimentally more viable in comparison to the same. 

\section{Acknowledgement}
We acknowledge computations performed using Armadillo~\cite{Sanderson12016, Sanderson1} on the cluster computing facility of the Harish-Chandra Research Institute, India. This research was supported in part by the `INFOSYS scholarship for senior students'. We also acknowledge partial support from the Department of Science and Technology, Government of India through the QuEST grant (grant number DST/ICPS/QUST/Theme-3/2019/120).

\onecolumngrid
\appendix
\section{Tables}
\label{appendix1}
\setcounter{table}{0}
\renewcommand{\thetable}{A \Roman{table}}
\begin{table*}[!htb]
    \centering
    \begin{tabular}{||c|c|c|c|c|c|c|c|c|c|c||}
    \hline
        \hspace{1mm} $\delta$ \hspace{1mm} & \hspace{1mm} 0 \hspace{1mm} & \hspace{1mm} 0.1 \hspace{1mm} & \hspace{1mm} 0.2 \hspace{1mm} & \hspace{1mm} \hspace{1mm} 0.3 \hspace{1mm} & \hspace{1mm} 0.4 \hspace{1mm} & \hspace{1mm} 0.5 \hspace{1mm} & \hspace{1mm} 0.6 \hspace{1mm} & \hspace{1mm} 0.7 \hspace{1mm} & \hspace{1mm} 0.8 \hspace{1mm}  & \hspace{1mm} 0.9 \hspace{1mm} \\ \hline \hline
        \hspace{1mm}N = 6 \hspace{1mm} & \hspace{1mm} 0.33334 \hspace{1mm} & \hspace{1mm} 0.32259 \hspace{1mm} & \hspace{1mm} 0.31250 \hspace{1mm} & \hspace{1mm} 0.30304 \hspace{1mm} &\hspace{1mm}  0.29412 \hspace{1mm} &\hspace{1mm}  0.28572 \hspace{1mm} & \hspace{1mm} 0.27778 \hspace{1mm} &\hspace{1mm}  0.27028 \hspace{1mm} & \hspace{1mm} 0.26316 \hspace{1mm} &\hspace{1mm}  0.25642 \hspace{1mm} \\ 
        \hspace{1mm} N = 8 \hspace{1mm} & \hspace{1mm} 0.32924 \hspace{1mm} & \hspace{1mm} 0.31856 \hspace{1mm} & \hspace{1mm} 0.30851 \hspace{1mm} & \hspace{1mm} 0.29906 \hspace{1mm} & \hspace{1mm} 0.29017\hspace{1mm} &\hspace{1mm} 0.28178 \hspace{1mm} & \hspace{1mm} 0.27387\hspace{1mm} & \hspace{1mm} 0.26640 \hspace{1mm} & \hspace{1mm} 0.25934 \hspace{1mm} & \hspace{1mm} 0.25265 \hspace{1mm} \\ 
        \hspace{1mm}N = 10 \hspace{1mm} &  \hspace{1mm}0.32736 \hspace{1mm} &  \hspace{1mm} 0.31668 \hspace{1mm} & \hspace{1mm} 0.30663 \hspace{1mm} & \hspace{1mm} 0.29717 \hspace{1mm} &  \hspace{1mm}0.28828 \hspace{1mm} & \hspace{1mm} 0.27989 \hspace{1mm} & \hspace{1mm} 0.27199 \hspace{1mm} &  \hspace{1mm} 0.26453 \hspace{1mm} & \hspace{1mm} 0.25749 \hspace{1mm} & \hspace{1mm} 0.25082 \hspace{1mm} \\ 
         \hspace{1mm} N = 12 \hspace{1mm} & \hspace{1mm} 0.32628 \hspace{1mm} & \hspace{1mm} 0.31560 \hspace{1mm} & \hspace{1mm} 0.30556 \hspace{1mm} & \hspace{1mm} 0.29611 \hspace{1mm} & \hspace{1mm} 0.28722 \hspace{1mm} & \hspace{1mm} 0.27884 \hspace{1mm} &  \hspace{1mm}0.27094\hspace{1mm} & \hspace{1mm} 0.26349 \hspace{1mm} &  \hspace{1mm} 0.25646 \hspace{1mm} & \hspace{1mm} 0.24980 \hspace{1mm}  \\ 
        N = 14 & 0.32562 & 0.31494 & 0.30490 & 0.29546 & 0.28657 & 0.2782 & 0.27031 & 0.26286 & 0.25583 & 0.24918 \\ 
        N = 16 & 0.32518 & 0.31450 & 0.30447 & 0.29503 & 0.28614 & 0.27778 & 0.26989 & 0.26245 & 0.25542 & 0.24877 \\ 
        N $\approx$ $\infty$ & 0.32347 & 0.31286 & 0.30284 & 0.29342 & 0.28464 & 0.27632 & 0.26862 & 0.26120 & 0.25404 &0.24732 \\ 
        \hspace{1mm} $\beta$ \hspace{1mm} & \hspace{1mm} 1.80 \hspace{1mm} & \hspace{1mm} 1.82 \hspace{1mm} & \hspace{1mm} 1.82 \hspace{1mm} &  \hspace{1mm} 1.83 \hspace{1mm} & \hspace{1mm} 1.88 \hspace{1mm} & \hspace{1mm} 1.90 \hspace{1mm} & \hspace{1mm} 2.00 \hspace{1mm} & \hspace{1mm} 2.01 \hspace{1mm} &  \hspace{1mm} 1.92 \hspace{1mm} & \hspace{1mm} 1.88 \hspace{1mm}  \\ \hline \hline 
     
    \end{tabular}
    \caption{Critical transition points $\alpha_c$ for different values of $N$ along with the extrapolated $N \rightarrow \infty$ limit for $\delta<1$ {using nearest neighbour concurrence of the subjacent state as the order parameter in the XXZ $J_1-J_2$ model}. The last row presents the finite-size scaling exponents at different $\delta$.}
    \label{delta0to1}
\end{table*}

\begin{table*}[h]
    \centering
    \begin{tabular}{||c|c|c|c|c|c|c|c|c|c|c||}
    \hline 
        \hspace{1mm} $\delta$ \hspace{1mm}& \hspace{1mm} 1 \hspace{1mm} & \hspace{1mm} 2 \hspace{1mm} & \hspace{1mm} 3 \hspace{1mm} & \hspace{1mm} 4 \hspace{1mm} & \hspace{1mm} 5 \hspace{1mm} & \hspace{1mm} 6 \hspace{1mm} & \hspace{1mm} 7 \hspace{1mm} & \hspace{1mm} 8 \hspace{1mm} & \hspace{1mm} 9 \hspace{1mm} & \hspace{1mm} 10 \hspace{1mm}  \\ \hline \hline
       \hspace{1mm} N = 6 \hspace{1mm} & \hspace{1mm} 0.25000 \hspace{1mm} & \hspace{1mm} 0.33334 \hspace{1mm} & \hspace{1mm} 0.37500 \hspace{1mm} & \hspace{1mm} 0.40000 \hspace{1mm} & \hspace{1mm} 0.41667 \hspace{1mm} &\hspace{1mm} 0.42858 \hspace{1mm} & \hspace{1mm} 0.4375 \hspace{1mm} & \hspace{1mm} 0.44445 \hspace{1mm} & \hspace{1mm} 0.45000 \hspace{1mm} & \hspace{1mm} 0.45455\hspace{1mm}  \\ 
       \hspace{1mm} N = 8 \hspace{1mm} & \hspace{1mm} 0.24630 \hspace{1mm} & \hspace{1mm} 0.32928 \hspace{1mm} & \hspace{1mm} 0.37127 \hspace{1mm} & \hspace{1mm} 0.39664 \hspace{1mm} & \hspace{1mm} 0.41363 \hspace{1mm} & \hspace{1mm} 0.42581 \hspace{1mm} & \hspace{1mm} 0.43497 \hspace{1mm} & \hspace{1mm} 0.44212 \hspace{1mm} & \hspace{1mm} 0.44784 \hspace{1mm} & \hspace{1mm} 0.45253 \hspace{1mm}  \\ 
       \hspace{1mm} N = 10 \hspace{1mm} & \hspace{1mm} 0.24449 \hspace{1mm} & \hspace{1mm} 0.32739 \hspace{1mm} & \hspace{1mm} 0.36956 \hspace{1mm} & \hspace{1mm} 0.39512 \hspace{1mm} & \hspace{1mm} 0.41228 \hspace{1mm} & \hspace{1mm} 0.4246 \hspace{1mm} &  \hspace{1mm} 0.43388 \hspace{1mm} & \hspace{1mm} 0.44111 \hspace{1mm} & \hspace{1mm} 0.44692 \hspace{1mm} & \hspace{1mm} 0.45168 \hspace{1mm}  \\ 
       \hspace{1mm} N = 12 \hspace{1mm} & \hspace{1mm} 0.24349 \hspace{1mm} & \hspace{1mm} 0.32634 \hspace{1mm} & \hspace{1mm} 0.36861 \hspace{1mm} & \hspace{1mm} 0.39428 \hspace{1mm} & \hspace{1mm} 0.41152 \hspace{1mm} & \hspace{1mm} 0.42392 \hspace{1mm} & \hspace{1mm} 0.43325 \hspace{1mm} & \hspace{1mm} 0.44054 \hspace{1mm} & \hspace{1mm} 0.44639 \hspace{1mm} & \hspace{1mm} 0.45120 \hspace{1mm} \\ 
       \hspace{1mm} N = 14 \hspace{1mm} & \hspace{1mm} 0.24288 \hspace{1mm} & \hspace{1mm} 0.32570 \hspace{1mm} & \hspace{1mm} 0.36803 \hspace{1mm} &\hspace{1mm} 0.39376 \hspace{1mm} & \hspace{1mm} 0.41106 \hspace{1mm} & \hspace{1mm} 0.42350 \hspace{1mm} & \hspace{1mm} 0.43288 \hspace{1mm} & \hspace{1mm} 0.44020 \hspace{1mm} &  \hspace{1mm}0.44608 \hspace{1mm} & \hspace{1mm} 0.45090 \hspace{1mm} \\ 
       \hspace{1mm} N = 16 \hspace{1mm} & \hspace{1mm} 0.24248 \hspace{1mm} & \hspace{1mm} 0.32528 \hspace{1mm} & \hspace{1mm} 0.36766 \hspace{1mm} & \hspace{1mm} 0.39342 \hspace{1mm} & \hspace{1mm} 0.41076 \hspace{1mm} & \hspace{1mm} 0.42323 \hspace{1mm} & \hspace{1mm} 0.43263 \hspace{1mm} & \hspace{1mm} 0.43997 \hspace{1mm} & \hspace{1mm} 0.44587 \hspace{1mm} & \hspace{1mm} 0.45071 \hspace{1mm}  \\ 
       \hspace{1mm} N $\approx$ $\infty$ \hspace{1mm} & \hspace{1mm} 0.24116 \hspace{1mm} & \hspace{1mm} 0.32382 \hspace{1mm} & \hspace{1mm} 0.36647 \hspace{1mm} & \hspace{1mm} 0.39226 \hspace{1mm} & \hspace{1mm} 0.40973 \hspace{1mm} & \hspace{1mm} 0.42228 \hspace{1mm} & \hspace{1mm} 0.43172 \hspace{1mm} & \hspace{1mm} 0.43916 \hspace{1mm} & \hspace{1mm} 0.44517 \hspace{1mm} & \hspace{1mm} 0.45002 \hspace{1mm}  \\ 
       \hspace{1mm} $\beta$ & 1.96 \hspace{1mm} &  \hspace{1mm} 1.91 \hspace{1mm} &\hspace{1mm} 2.00 \hspace{1mm} &\hspace{1mm} 1.93 \hspace{1mm} & \hspace{1mm} 1.95 \hspace{1mm} & \hspace{1mm} 1.93 \hspace{1mm} & \hspace{1mm} 1.89 \hspace{1mm} & \hspace{1mm} 1.92 \hspace{1mm} & \hspace{1mm} 1.97 \hspace{1mm} & \hspace{1mm} 1.92 \hspace{1mm}  \\ \hline \hline
       
    \end{tabular}
    
    \caption{Critical transition points $\alpha_c$ for different values of $N$ along with the extrapolated $N \rightarrow \infty$ limit for $\delta \ge 1$ using nearest neighbour concurrence of the subjacent state as the order parameter in the XXZ $J_1-J_2$ model. The last row presents the finite size scaling exponents at different $\delta$.}
    \label{delta1to10}
\end{table*}
\begin{table*}[h]
\centering
\begin{tabular}{|| c || c  c | c  c | c  c | c  c | c  c ||} 
\hline
\hspace{2mm}\multirow{2}{*}{$\sigma$}\hspace{2mm} & \multicolumn{2}{c}{$N = 6$}\hspace{1mm} \vline & \multicolumn{2}{c}{$N = 8$}\hspace{1mm} \vline & \multicolumn{2}{c}{$N = 10$} \hspace{1mm} \vline &  \multicolumn{2}{c}{$N = 12$} \hspace{1mm} \vline &  \multicolumn{2}{c}{$N = 14$} \hspace{1mm} \vline \\ [0.5ex] 
 & \hspace{1mm} $\alpha_C$ \hspace{1mm} & \hspace{1mm} error \hspace{1mm} & \hspace{1mm} $\alpha_C$ \hspace{1mm} & \hspace{1mm} error \hspace{1mm} & \hspace{1mm} $\alpha_C$ \hspace{1mm} & \hspace{1mm} error \hspace{1mm} & \hspace{1mm}$\alpha_C$ \hspace{1mm} & \hspace{1mm} error \hspace{1mm} & \hspace{1mm} $\alpha_C$ \hspace{1mm} & \hspace{1mm} error \hspace{1mm} \\ [0.5ex]
 \hline\hline
0.05 & \hspace{2mm} $0.25434$ \hspace{1mm} & \hspace{1mm} $0.0081$ \hspace{1mm} & \hspace{1mm} $0.24881$ \hspace{1mm} & \hspace{1mm} $0.0044$ \hspace{1mm} & \hspace{1mm} $0.24608$ \hspace{1mm} & \hspace{1mm} $0.0036$ \hspace{1mm} & \hspace{1mm}$0.24820$ \hspace{1mm} & \hspace{1mm} $0.0028$ \hspace{1mm} & \hspace{1mm} $0.24834$ \hspace{1mm} & \hspace{1mm} $0.0023$ \hspace{1mm} \\ 
0.08 & \hspace{2mm} $0.25521$ \hspace{1mm} & \hspace{1mm} $0.0068$ \hspace{1mm} & \hspace{1mm} $0.25038$ \hspace{1mm} & \hspace{1mm} $0.00085$ \hspace{1mm} & \hspace{1mm} $0.24750$ \hspace{1mm} & \hspace{1mm} $0.0024$ \hspace{1mm} & \hspace{1mm}$0.25002$ \hspace{1mm} & \hspace{1mm} $0.0013$ \hspace{1mm} & \hspace{1mm} $0.24968$ \hspace{1mm} & \hspace{1mm} $0.00074$ \hspace{1mm} \\ 
0.10 & \hspace{2mm} $0.25567$ \hspace{1mm} & \hspace{1mm} $0.0061$ \hspace{1mm} & \hspace{1mm} $0.25154$ \hspace{1mm} & \hspace{1mm} $0.0011$ \hspace{1mm} & \hspace{1mm} $0.25415$ \hspace{1mm} & \hspace{1mm} $0.0020$ \hspace{1mm} & \hspace{1mm}$0.25416$ \hspace{1mm} & \hspace{1mm} $0.0013$ \hspace{1mm} & \hspace{1mm} $0.25629$ \hspace{1mm} & \hspace{1mm} $0.00057$ \hspace{1mm} \\ 
0.13 & \hspace{2mm} $0.25975$ \hspace{1mm} & \hspace{1mm} $0.0030$ \hspace{1mm} & \hspace{1mm} $0.25804$ \hspace{1mm} & \hspace{1mm} $0.0010$ \hspace{1mm} & \hspace{1mm} $0.25761$ \hspace{1mm} & \hspace{1mm} $0.00053$ \hspace{1mm} & \hspace{1mm}$0.26736$ \hspace{1mm} & \hspace{1mm} $0.00096$ \hspace{1mm} & \hspace{1mm} $0.25547$ \hspace{1mm} & \hspace{1mm} $0.00085$ \hspace{1mm} \\ 
0.15 & \hspace{2mm} $0.26095$ \hspace{1mm} & \hspace{1mm} $0.0026$ \hspace{1mm} & \hspace{1mm} $0.26053$ \hspace{1mm} & \hspace{1mm} $0.00045$ \hspace{1mm} & \hspace{1mm} $0.26215$ \hspace{1mm} & \hspace{1mm} $0.0012$ \hspace{1mm} & \hspace{1mm}$0.25951$ \hspace{1mm} & \hspace{1mm} $0.00075$ \hspace{1mm} & \hspace{1mm} $0.27500$ \hspace{1mm} & \hspace{1mm} $0.00078$ \hspace{1mm} \\ 
0.18 & \hspace{2mm} $0.26237$ \hspace{1mm} & \hspace{1mm} $0.0019$ \hspace{1mm} & \hspace{1mm} $0.26618$ \hspace{1mm} & \hspace{1mm} $0.00079$ \hspace{1mm} & \hspace{1mm} $0.26659$ \hspace{1mm} & \hspace{1mm} $0.00057$ \hspace{1mm} & \hspace{1mm}$0.27478$ \hspace{1mm} & \hspace{1mm} $0.00078$ \hspace{1mm} & \hspace{1mm} $0.23801$ \hspace{1mm} & \hspace{1mm} $0.00065$ \hspace{1mm} \\ 
0.20 & \hspace{2mm} $0.26442$ \hspace{1mm} & \hspace{1mm} $0.0018$ \hspace{1mm} & \hspace{1mm} $0.26821$ \hspace{1mm} & \hspace{1mm} $0.00085$ \hspace{1mm} & \hspace{1mm} $0.27002$ \hspace{1mm} & \hspace{1mm} $0.00089$ \hspace{1mm} & \hspace{1mm}$0.27985$ \hspace{1mm} & \hspace{1mm} $0.00078$ \hspace{1mm} & \hspace{1mm} $0.27977$ \hspace{1mm} & \hspace{1mm} $0.00042$ \hspace{1mm} \\ 
 \hline \hline 
\end{tabular}
\caption{The values of $\alpha_c$ for different sizes of the system (N) and different $\sigma$ for the mixing probability $p = 0.2689$, in the $J_1-J_2$ Heisenberg chain with glassy anisotropy. }
\label{table:disorder}
\end{table*}

\begin{table*}[h]
\centering
\begin{tabular}{|| c || c  c | c  c | c  c | c  c | c  c ||} 
\hline
\hspace{2mm}\multirow{2}{*}{$\sigma$}\hspace{2mm} & \multicolumn{2}{c}{$N = 6$}\hspace{1mm} \vline & \multicolumn{2}{c}{$N = 8$}\hspace{1mm} \vline & \multicolumn{2}{c}{$N = 10$} \hspace{1mm} \vline &  \multicolumn{2}{c}{$N = 12$} \hspace{1mm} \vline &  \multicolumn{2}{c}{$N = 14$} \hspace{1mm} \vline \\ [0.5ex] 
& \hspace{1mm} $\alpha_C$ \hspace{1mm} & \hspace{1mm} error \hspace{1mm} & \hspace{1mm} $\alpha_C$ \hspace{1mm} & \hspace{1mm} error \hspace{1mm} & \hspace{1mm} $\alpha_C$ \hspace{1mm} & \hspace{1mm} error \hspace{1mm} & \hspace{1mm}$\alpha_C$ \hspace{1mm} & \hspace{1mm} error \hspace{1mm} & \hspace{1mm} $\alpha_C$ \hspace{1mm} & \hspace{1mm} error \hspace{1mm} \\ [0.5ex]
 \hline\hline
0.05 & \hspace{2mm} $0.25201$ \hspace{1mm} & \hspace{1mm} $0.0046$ \hspace{1mm} & \hspace{1mm} $0.24879$ \hspace{1mm} & \hspace{1mm} $0.0020$ \hspace{1mm} & \hspace{1mm} $0.24799$ \hspace{1mm} & \hspace{1mm} $0.0016$ \hspace{1mm} & \hspace{1mm}$0.24826$ \hspace{1mm} & \hspace{1mm} $0.0015$ \hspace{1mm} & \hspace{1mm} $0.24465$ \hspace{1mm} & \hspace{1mm} $0.0014$ \hspace{1mm} \\ 
0.08 & \hspace{2mm} $0.25353$ \hspace{1mm} & \hspace{1mm} $0.0043$ \hspace{1mm} & \hspace{1mm} $0.25034$ \hspace{1mm} & \hspace{1mm} $0.00071$ \hspace{1mm} & \hspace{1mm} $0.25225$ \hspace{1mm} & \hspace{1mm} $0.0012$ \hspace{1mm} & \hspace{1mm}$0.25044$ \hspace{1mm} & \hspace{1mm} $0.000033$ \hspace{1mm} & \hspace{1mm} $0.25357$ \hspace{1mm} & \hspace{1mm} $0.00078$ \hspace{1mm} \\ 
0.10 & \hspace{2mm} $0.25567$ \hspace{1mm} & \hspace{1mm} $0.0031$ \hspace{1mm} & \hspace{1mm} $0.25201$ \hspace{1mm} & \hspace{1mm} $0.0000014$ \hspace{1mm} & \hspace{1mm} $0.25557$ \hspace{1mm} & \hspace{1mm} $0.00099$ \hspace{1mm} & \hspace{1mm}$0.25363$ \hspace{1mm} & \hspace{1mm} $0.00048$ \hspace{1mm} & \hspace{1mm} $0.26879$ \hspace{1mm} & \hspace{1mm} $0.00081$ \hspace{1mm} \\ 
0.13 & \hspace{2mm} $0.25860$ \hspace{1mm} & \hspace{1mm} $0.0055$ \hspace{1mm} & \hspace{1mm} $0.25701$ \hspace{1mm} & \hspace{1mm} $0.0024$ \hspace{1mm} & \hspace{1mm} $0.25590$ \hspace{1mm} & \hspace{1mm} $0.0012$ \hspace{1mm} & \hspace{1mm}$0.25849$ \hspace{1mm} & \hspace{1mm} $0.00081$ \hspace{1mm} & \hspace{1mm} $0.26155$ \hspace{1mm} & \hspace{1mm} $0.00092$ \hspace{1mm} \\ 
0.15 & \hspace{2mm} $0.25943$ \hspace{1mm} & \hspace{1mm} $0.0044$ \hspace{1mm} & \hspace{1mm} $0.25749$ \hspace{1mm} & \hspace{1mm} $0.0023$ \hspace{1mm} & \hspace{1mm} $0.25878$ \hspace{1mm} & \hspace{1mm} $0.0010$ \hspace{1mm} & \hspace{1mm}$0.26269$ \hspace{1mm} & \hspace{1mm} $0.00070$ \hspace{1mm} & \hspace{1mm} $0.26787$ \hspace{1mm} & \hspace{1mm} $0.00094$ \hspace{1mm} \\ 
0.18 & \hspace{2mm} $0.26058$ \hspace{1mm} & \hspace{1mm} $0.0034$ \hspace{1mm} & \hspace{1mm} $0.26094$ \hspace{1mm} & \hspace{1mm} $0.0011$ \hspace{1mm} & \hspace{1mm} $0.26262$ \hspace{1mm} & \hspace{1mm} $0.0011$ \hspace{1mm} & \hspace{1mm}$0.26387$ \hspace{1mm} & \hspace{1mm} $0.00082$ \hspace{1mm} & \hspace{1mm} $0.27721$ \hspace{1mm} & \hspace{1mm} $0.00098$ \hspace{1mm} \\ 
0.20 & \hspace{2mm} $0.26540$ \hspace{1mm} & \hspace{1mm} $0.0037$ \hspace{1mm} & \hspace{1mm} $0.26238$ \hspace{1mm} & \hspace{1mm} $0.00095$ \hspace{1mm} & \hspace{1mm} $0.26604$ \hspace{1mm} & \hspace{1mm} $0.00062$ \hspace{1mm} & \hspace{1mm}$0.26943$ \hspace{1mm} & \hspace{1mm} $0.00064$ \hspace{1mm} & \hspace{1mm} $0.28565$ \hspace{1mm} & \hspace{1mm} $0.00079$ \hspace{1mm} \\ 
 \hline \hline 
\end{tabular}
\caption{The values of $\alpha_c$ for different sizes of the system (N) and different $\sigma$ for the mixing probability $p = 0.1344$, in the $J_1-J_2$ Heisenberg chain with glassy anisotropy.}
\label{table:disorderp13}
\end{table*}


\twocolumngrid
\bibliography{References}

\end{document}